# Novel distance-based masking and adaptive $\alpha$-shape methods for CNN-ready reconstruction of arbitrary 2D CFD flow domains

Mehran Sharifi, Gorka S. Larraona, and Alejandro Rivas

*Universidad de Navarra, TECNUN, Department of Mechanical Engineering and Materials, Manuel de Lardizabal 13, 20018 Donostia / San Sebastián, Spain.*

**Abstract**

Interpolating scattered CFD datasets onto a uniform Cartesian grid can distort the true geometry, producing a convex-hull type envelope and activating nonphysical regions. This work presents a reconstruction framework that recovers physically consistent masks before exporting CNN-ready fields. It introduces two novel strategies, distance-based masking and an adaptive $\alpha$-shape formulation that normalizes $\alpha$ using local data resolution, and evaluates them against classical $\alpha$-shape boundary recovery. A quantitative, topology-aware metric suite is introduced to assess retention, suppression of unsupported regions, overlap consistency, and connectivity. The novel distance-based method is robust across the geometries considered under the same threshold rule, with $\tau$ set to the minimum CFD grid spacing, and achieves $500\times - 800\times$ speedups over classical $\alpha$-shapes. The adaptive $\alpha$-shape remains stable when its control parameter is set to 1 and is $1.7\times - 2.6\times$ faster than the classical variant, which requires geometry-specific $\alpha$ tuning. A lightweight boundary inflation post-process using a minimal dilation further improves retention by up to 2.96% with negligible unsupported activation ($\leq 0.08\%$). Overall, the distance-based method is recommended as the default due to its accuracy, stability, minimal tuning, and low cost, while the adaptive $\alpha$-shape is a strong alternative when grid-spacing information for threshold selection is unavailable. A companion web application operationalizes the workflow end to end, enabling 2D ASCII dataset upload, parameter tuning, mask and boundary generation, and export of CNN-ready outputs.

# Nomenclature

**Latin symbols**

| | | |
|---|---|---|
| $A_{0,1,2}$ | : | Triangle area, $[m^2]$ |
| AVF | : | Active volume fraction criterion, $[-]$ |
| $B$ | : | Structuring element, $[-]$ |
| $\mathcal{B}$ | : | Axis-aligned bounding box, $[-]$ |
| $\boldsymbol{c_\sigma}$ | : | Circumcenter, $[m]$ |
| $C$ | : | Geometric complex, $[-]$ |
| CC | : | Connected component criterion, $[-]$ |
| $d$ | : | Spatial dimension, $[-]$ |
| $d^{NN}$ | : | Nearest neighbor distance, $[m]$ |
| $D$ | : | Euclidean distance map, $[m]$ |
| $D(\boldsymbol{x_j'})$ | : | Distance field value, $[m]$ |
| $e$ | : | Delaunay edge length, $[m]$ |
| $\mathcal{E}$ | : | Set of Delaunay edges, $[-]$ |
| EF | : | Boundary elements, $[m]$ |
| $\boldsymbol{f}$ | : | Sampled field quantity, $[*]$ |
| $\mathcal{G}$ | : | Ghost voxel set, $[-]$ |
| $G$ | : | Set of structured grid nodes, $[-]$ |
| GF | : | Ghost fraction criterion, $[-]$ |
| IoP | : | Intersection over union criterion, $[-]$ |
| $\mathcal{I}$ | : | Intersection (or overlap) mask, $[-]$ |
| $\boldsymbol{j}$ | : | Multi-index vector, $[-]$ |
| $L$ | : | Section lengths, $[m]$ |
| $L_{ac}$ | : | Axial chord, $[m]$ |
| $L_c$ | : | Chord, $[m]$ |
| $\mathcal{L}$ | : | Structured grid index set, $[-]$ |
| $M$ | : | In-domain set, $[-]$ |
| $\boldsymbol{n}$ | : | CNN grid resolution vector, $[-]$ |
| $n_{ef}$ | : | Element occurrence count, $[-]$ |
| $N$ | : | Number of data samples, $[-]$ |
| $N_e$ | : | Number of unique edges, $[-]$ |
| $\boldsymbol{p'}$ | : | Query point, $[m]$ |
| PR | : | Point recall criterion, $[-]$ |
| Pre | : | Precision criterion, $[-]$ |
| $\mathcal{P}$ | : | Set of scattered data samples, $[-]$ |
| $q$ | : | Field dimension, $[-]$ |
| $r_0$ | : | Reference radius, $[-]$ |
| $r_c$ | : | Chebyshev radius, $[-]$ |
| $r_\sigma$ | : | Circumradius, $[m]$ |
| $R$ | : | Section widths, $[m]$ |
| Rec | : | Recall criterion, $[-]$ |
| $s$ | : | Structuring element size, $[-]$ |
| $S$ | : | Circumball, $[m^2]$ |
| Sh | : | $\alpha$-shape of point set, $[m^2]$ |
| $\mathcal{U}$ | : | Union mask, $[-]$ |
| $\boldsymbol{x}$ | : | Data-point coordinate vector, $[m]$ |
| $\boldsymbol{x'}$ | : | Grid-node coordinate vector, $[m]$ |

**Greek symbols**

| | | |
|---|---|---|
| $\alpha$ | : | $\alpha$-shape parameter, $[m]$ |
| $\alpha_a$ | : | Adaptive $\alpha$-shape parameter, $[m^{-1}]$ |
| $\beta$ | : | Control parameter, $[-]$ |

**Greek symbols (continue)**

| | | |
|---|---|---|
| $\gamma$ | : | Interpolation weight, $[-]$ |
| $\delta$ | : | Empirical scaling coefficient, $[-]$ |
| $\eta$ | : | Mask expansion factor, $[-]$ |
| $\lambda$ | : | Barycentric coordinate, $[-]$ |
| $\pi(\boldsymbol{x_i})$ | : | Grid-index mapping, $[-]$ |
| $\rho$ | : | Neighborhood scaling factor, $[-]$ |
| $\sigma$ | : | Delaunay Simplex, $[m^2]$ |
| $\tau$ | : | Distance threshold, $[m]$ |
| $\varphi$ | : | Exit blade angle, $[°]$ |
| $2\phi$ | : | Bifurcation angle, $[°]$ |
| $\chi$ | : | Binary indicator function, $[-]$ |
| $\omega$ | : | Inlet blade angle, $[°]$ |

**Subscripts**

| | | |
|---|---|---|
| $\alpha$ | : | $\alpha$-shape method |
| $av$ | : | Active voxel |
| $a\alpha$ | : | Adaptive $\alpha$-shape method |
| $c$ | : | Connected compartment |
| $db$ | : | Distance-based method |
| $eb$ | : | External boundary element |
| $ef$ | : | Arbitrary boundary element |
| $f$ | : | Final value |
| $i$ | : | Inlet-section value |
| $in$ | : | Inner value |
| $k$ | : | Arbitrary reconstruction method |
| $m$ | : | Main-section (middle) value |
| $max$ | : | Maximum value |
| $min$ | : | Minimum value |
| $o$ | : | Outlet-section value |
| $opt$ | : | Optimal value |
| $out$ | : | Outer value |

**Superscripts**

| | | |
|---|---|---|
| $\bar{\phantom{x}}$ | : | Mean value |
| $*$ | : | Interpolated value |

**Special Operators**

| | | |
|---|---|---|
| $\odot$ | : | Hadamard product |
| $\ominus$ | : | Binary erosion |
| $\oplus$ | : | Binary dilation |
| $\cup$ | : | Union |
| $\cap$ | : | Intersection |
| $\Delta$ | : | Grid spacing |
| $\mathbf{1}(\cdot)$ | : | Counting |
| $|\cdot|$ | : | Set cardinality |
| $|\cdot|_{(a-v)}$ | : | Simplex measure |
| $\|\cdot\|_2$ | : | Euclidean norm |
| $\text{conv}(.)$ | : | Convex hull |
| $\det(.)$ | : | Matrix determinant |
| $\text{Del}(.)$ | : | Delaunay tessellation |
| $\max(.)$ | : | Maximum |

| **Abbreviations** | | |
|---|---|---|
| 2D | : | Two-Dimensional |
| 3D | : | Three-Dimensional |
| AGB | : | Above-Ground Biomass |
| APT | : | Atom Probe Tomography |
| BFS | : | Breadth-First Search |
| CAA | : | Cluster And Adjustment |
| CFD | : | Computational Fluid Dynamics |
| CNN | : | Convolutional Neural Network |
| CT | : | Computed Tomography |
| DP | : | Douglas-Peucker |
| GAT | : | Graph Attention Network |
| GCN | : | Graph Convolutional Network |
| GeoGen | : | Geometry-aware Generative |
| GPU | : | Graphics Processing Unit |
| GraphSAGE | : | Graph SAmple and aggreGatE |
| KD | : | K-Dimensional |
| KPConv | : | Kernel Point Convolution |
| LiDAR | : | Light Detection And Ranging |
| MRF | : | Markov Random Field |
| MRI | : | Magnetic Resonance Imaging |
| NDF | : | Neural Distance Field |
| NN | : | Nearest Neighbor |
| PFEM | : | Particle Finite Element Method |
| RGB | : | Red, Green, and Blue |
| SDF | : | Signed Distance Field |
| TIN | : | Triangulated Irregular Network |
| UAV | : | Unmanned Aerial Vehicle |
| UDF | : | Unsigned Distance Field |

## 1. Introduction

Convolutional Neural Networks (CNNs) have emerged as one of the most powerful architectures for extracting hierarchical and spatially correlated features from structured data [1-3]. Originally developed for computer vision [4, 5], they have proven highly effective in learning multi-scale representations from complex scientific and engineering datasets [6-8]. Their strength lies in the convolutional operation, which employs local receptive fields and weight sharing to capture spatial dependencies while ensuring translational invariance and computational efficiency. By integrating convolutional, pooling, normalization, and activation layers, CNNs progressively learn both low- and high-level abstractions of physical fields and patterns. When trained on properly structured and normalized datasets, CNNs can act as fast, data-driven surrogates for computationally expensive simulations and as versatile tools for feature extraction, dimensionality reduction, and pattern recognition [9-12]. Their ability to approximate nonlinear mappings between boundary conditions and physical responses enables near-real-time prediction and optimization across many domains. In fluid dynamics and heat transfer, CNNs have been used to reconstruct velocity [13, 14], pressure [15, 16], and temperature [17, 18] fields, predict turbulent structures [19, 20], and model thermo-fluid phenomena such as convection, boundary-layer development, and flow separation [21-23]. They accelerate high-fidelity simulations and support the design of efficient thermal systems, including heat exchangers, energy storage units, and phase-change processes [24-26]. Beyond thermo-fluids, CNNs play a major role in materials science for microstructural analysis and property prediction [27, 28], medical imaging for segmentation and diagnosis [29, 30], and geosciences for weather forecasting, seismic interpretation, and groundwater modeling [31-33]. By efficiently capturing spatial dependencies and preserving geometric fidelity, CNNs bridge data-driven inference with physical interpretability, motivating the development of preprocessing and domain-reconstruction frameworks for generating CNN-ready structured datasets.

The performance of CNNs fundamentally depends on the availability of structured and grid-aligned datasets, as convolution operations assume spatial regularity with uniformly distributed neighboring points. Such representations enable consistent kernel application, efficient computation, and progressive feature extraction across multiple scales. In contrast, most scientific and engineering datasets are unstructured or scattered, obtained from irregular meshes, experiments, or point clouds with missing regions [34, 35]. To address this, point-based neural networks such as PointNet [36], PointNet++ [37], PointConv [38], and Kernel Point Convolution (KPConv) [39], along with graph-based architectures such as Graph Convolutional Network (GCN) [40], Graph SAmple and aggreGatE (GraphSAGE) [41], ChebNet [42], Graph Attention Network (GAT) [43], and MeshCNN [44], have been developed to process irregular or non-Euclidean data directly. These methods effectively preserve geometric and topological relationships without requiring data regularization. However, they often entail high computational cost, complex neighborhood construction, and limited efficiency in capturing local spatial correlations for large-scale multidimensional fields [45, 46]. In comparison, CNNs leverage structured grids with well-defined receptive fields, offering efficient convolution operations, spatial translation invariance, and straightforward parallelization on modern Graphics Processing Units (GPUs). Their scalability, computational efficiency, and training stability make them particularly suitable for large scientific and engineering datasets. Consequently, transforming unstructured or scattered data into structured, grid-aligned representations through data reconstruction and domain regularization remains a practical and powerful approach [5, 47]. Yet, accurately recovering domain boundaries, especially near irregular or disconnected regions, remains a central challenge for maintaining geometric fidelity and physical consistency. This difficulty arises because uniform-grid interpolation inherently distorts the geometry, often creating a convex envelope that includes spurious regions. This necessitates reconstruction methods capable of resolving the true concave physical boundaries from these distorted representations [48].

Distance-based boundary recovery methods, often implemented through thresholding or level-set formulations, provide a simple yet highly effective strategy for reconstructing geometric domains from scattered scientific or engineering datasets. The core idea is to compute a distance field between each point on a structured grid and the nearest scattered sample, typically using efficient Nearest-Neighbor (NN) searches such as K-Dimensional (KD) trees [49], and then classify grid nodes based on a distance threshold. Points within the threshold are identified as interior, while the rest are masked as exterior. Because the Signed Distance Field (SDF) inherently captures geometric proximity, the method can represent complex shapes, disconnected components, and highly irregular boundaries without explicit meshing [50, 51]. Table 1 offers a structured overview of relevant studies that have applied this approach across different scientific and engineering contexts.

**Table 1** Overview of distance-based boundary recovery techniques for geometric domain reconstruction.

| Author(s) | Application Domain | Purpose of Study | Method Type | Key Contribution |
|---|---|---|---|---|
| Chang et al. [52] | Geometry processing | Efficient local SDF generation for polygonal meshes | Local SDF, voxelization | Volumetric SDF computed only in a narrow band; voxel–triangle reduction via triangle lists; scalable to large meshes; avoided GPU memory limits. |
| Calakli and Taubin [53] | Implicit surface reconstruction | Closed-surface recovery from oriented point samples | Variational SDF reconstruction, dual marching cubes | Developed smooth SDF formulation; normals integrated directly; single variational solve; adaptive octree; high-quality crack-free meshes; simpler than Poisson-based methods. |
| Sharma and Agrawal [54] | Geometry processing, cross-sectional reconstruction | Recover smooth Three-Dimensional (3D) surfaces | Globally consistent SDF on planes, Hermite mean-value interpolation | Reconstructed smooth closed surfaces from sparse, unorganized cross-sections; improved robustness with better normals; identified staircase artifacts due to slice-aligned normals; outlined extensions for linear cross-sections. |
| Tang and Feng [55] | Shape modeling | Multi-scale implicit reconstruction | Adaptive SDF, hierarchical B-splines | Built curvature-adaptive SDF; octree grid captured geometric complexity at proper scales; hierarchical B-splines produced sparse well-conditioned system; produced detailed, crack-free adaptive meshes with strong runtime/memory performance. |
| Krayer and Müller [56] | Graphics processing | Robust SDF computation | Fully GPU-resident unsigned distance, winding-number sign using ray map | High-performance SDF generation; ray-map reduces ray–triangle tests; robust to holes, missing regions, and self-intersections. |
| Chibane et al. [57] | Neural implicit reconstruction | Learnable high-resolution representation for open surfaces | Neural Distance Field (NDF) | NDF predicted unsigned distance fields without requiring closed surfaces; reconstructed open and internally complex shapes; supported normals and sphere-tracing rendering; enabled multi-target regression. |
| Jiang et al. [58] | Computer vision | Optimize 3D geometry from images | Differentiable SDF-based rendering, multi-resolution optimization | SDF enabled flexible topology and watertight surfaces; reconstructed detailed multi-view 3D models; integrated with neural networks for learning without 3D ground truth. |
| Lin et al. [59] | Computer vision | Learn 3D shapes using only Two-Dimensional (2D) silhouettes | Implicit SDF, silhouette-only differentiable rendering | Learned from a single viewpoint per object; handled complex topologies; outperformed prior silhouette-based single-view methods on synthetic and real datasets. |
| Chen et al. [60] | Building modeling | Generate compact watertight building meshes | SDF, adaptive cell complex, Markov Random Field (MRF) optimization | Built an adaptive polyhedral cell complex; polygon extraction from SDFs without iso-surfacing; learned a continuous implicit occupancy field for interior/exterior classification. |
| Long et al. [61] | Neural implicit reconstruction | Recover surfaces with | Unsigned Distance Field (UDF), SDF, | Used UDFs to reconstruct open/non-closed surfaces; introduced density |

| Author(s) | Application Domain | Purpose of Study | Method Type | Key Contribution |
|---|---|---|---|---|
| | | flexible topology | modified volume rendering | function tied to UDF; competitive with SDF methods for closed shapes. |
| Shim et al. [62] | Implicit shape synthesis | Generate high-resolution 3D shapes | Coarse SDF, super-resolution diffusion | Two-stage modeling enabled detailed SDF generation; patch-wise super-resolution managed memory; extended naturally to multi-category generation and shape completion. |
| Esposito [63] | Medical 3D reconstruction | Reconstruct complex thin structures and anatomy | NDF, CrossSDF, Geometry-aware Generative (GeoGen) | Reconstructed thin vessels and preserved topology; enabled surgical planning and reduced Computed Tomography (CT) dose; general SDF framework for sparse 3D reconstruction. |
| Li et al. [64] | Indoor neural reconstruction | Recover indoor surfaces with low texture | Plane-guided SDF regularization, adaptive plane weighting | Introduced pseudo-plane detection with adaptive weights; SDF plane regularization; improved reconstruction on low-texture indoor scenes. |
| Hubert-Brierre et al. [65] | Implicit surface rendering | Accelerate SDF evaluation in construction trees | SDF, optimization nodes, proxy nodes, normal-warping | Accelerated SDF evaluation in complex construction trees; preserved Lipschitz/conservative SDF behavior; achieved major speed-ups in sphere-tracing and polygonization with lightweight normal-warping. |

Computational geometry offers another class of boundary recovery methods that infer domain shape from local relationships in scattered datasets [66]. A widely used approach is the $\alpha$-shape (or $\alpha$-complex), which forms a flexible non-convex boundary by filtering the Delaunay triangulation and retaining only simplices whose circumsphere radii fall below a specified $\alpha$, thereby capturing concavities, cavities, and complex topology [67, 68]. However, a single global $\alpha$ can over-smooth the boundary or fragment it when sampling density varies. Adaptive $\alpha$-shape variants address this by scaling $\alpha$ locally to preserve thin features, narrow gaps, and heterogeneous regions, yielding more robust CNN-ready structured masks from irregular data [69, 70]. $\alpha$-shapes and their adaptive extensions have been applied across LiDAR modeling, medical imaging, geophysics, forestry, and particle simulations, as summarized in Table 2.

**Table 2** Overview of $\alpha$-shape and adaptive $\alpha$-shape techniques in boundary recovery and geometric reconstruction.

| Author(s) | Application Domain | Purpose of Study | Method Type | Key Contribution |
|---|---|---|---|---|
| Wei [71] | LiDAR urban modeling | Building-boundary extraction | $\alpha$-shape, polygon filtering, boundary regularization | Used $\alpha$-shape to captured concave/convex outlines without Triangulated Irregular Network (TIN); robust noise removal; refined, consistent footprints. |
| Li et al. [72] | LiDAR buildings | Automated footprint delineation | Dual-threshold $\alpha$-shape, Breadth-First | Two $\alpha$ values improved detail/completeness balance; BFS ensured closed boundaries; least- |

| | | | Search (BFS), least-squares refinement | squares smoothing outperformed Douglas-Peucker (DP) algorithm; improved geometric fidelity. |
|---|---|---|---|---|
| Al-Tamimi et al. [73] | Medical imaging, Magnetic Resonance Imaging (MRI) | Brain-tumor volume estimation | Active contours, 3D $\alpha$-shape | Improved volumetric accuracy by utilizing $\alpha$-shape; interactive MATLAB tool; better precision than standard neuroradiological methods. |
| Varytimidis et al. [74] | Image processing, feature detection | Extract robust edge-based features | Anisotropic $\alpha$-shape, triangulation filtering | Developed $\alpha$-filtration–based feature detection robust to scale, rotation, affine transformations, illumination changes, and blur. |
| Gardiner et al. [75] | Biological morphology, $\mu$CT shape analysis | Measure 3D shape complexity of irregular structures | Multi-scale $\alpha$-shape | Used $\alpha$-shape to capture morphological complexity; optimal $\alpha$ chosen when $\alpha$-shape volume matched voxel volume; captured concavities better than classical metrics. |
| Bonneau et al. [76] | Geohazards, rockfall modeling | Rockfall block volume estimation | Power crust, convex hull, iterative $\alpha$-shape | Showed watertightness critical; $\alpha$-shapes sensitive to $\alpha$-radius and density; Convex Hull overestimates; Power Crust dense but watertight. |
| Still et al. [77] | Atom Probe Tomography (APT) | Improve cluster geometry and composition estimation | $\alpha$-Shape | Built $\alpha$-shape meshes around clusters to compute volume, area, and composition; handled concave clusters better than ellipsoidal fits; highlighted limitation of fixed $\alpha$ and clustering dependence. |
| Akdim et al. [78] | General 3D reconstruction | Compare reconstruction methods for scanned objects | Ball pivoting, $\alpha$-shape, model-based methods | Found Ball Pivoting gave highest-quality meshes, $\alpha$-shape provided fastest reconstruction; classified methods into combinatorial vs. model-based. |
| Zhao and Hu [79] | Tunnel deviations, civil engineering | Reconstruct tunnel geometries and visualize over/under-excavation | Low-parameter $\alpha$-shape, voxel filtering, KD tree rendering | Proposed a fast $\alpha$-shape reconstruction pipeline with voxel simplification and Red, Green, and Blue (RGB) deviation mapping; enabled high-speed visualization of tunnel excavation errors. |
| Chen and Yao [80] | LiDAR buildings | Boundary extraction and artifact reduction | $\alpha$-shape, 2D projection, recursive Gaussian smoothing, DP | Controlled boundary with $\alpha$-shape; improved boundary extraction accuracy by 43.79% vs. Cluster And Adjustment (CAA) method; reduced sawtooth artifacts; reduced complexity from $O(n^2)$ to $O(n)$. |
| Ismail et al. [81] | Building digital twins | Evaluate reconstruction methods for building point clouds | Poisson, Ball Pivoting, $\alpha$-shape, smoothing refinement | Showed Poisson gives watertight but over-triangulated meshes; Ball Pivoting preserves edges but leaves holes; $\alpha$-shapes sensitive to sampling; recommended hybrid methods. |
| Jang and Kang [82] | Unmanned Aerial Vehicle (UAV) monitoring, riparian vegetation | Vegetation-volume estimation | $\alpha$-shape | $\alpha \approx 0.75$ gave best accuracy; improved river-management assessment under climate instability. |
| Fernández et al. [83] | Particle Finite Element Method (PFEM), free-surface flows | Improve PFEM remeshing | $\alpha$-shape, level-set | Replaced $\alpha$-shape boundary detection with level-set method for smoother free surfaces; reduced mass loss; |

| | | | | improved droplet meshing; controlled element formation. |
|---|---|---|---|---|
| Wang et al. [84] | Forestry, tree reconstruction | Generate complete 3D tree models | Morphological segmentation, $\alpha$-shape, trunk fusion | $\alpha$-shape constrained missing-branch reconstruction; outperformed state-of-the-art methods; robust to noise, occlusion, and gaps. |
| Wang et al. [85] | Forestry, biomass estimation | Improve Above-Ground Biomass (AGB) estimation | $\alpha$-shape, voxelization for crown volume | Combined $\alpha$-shape and voxel crown volume best matched "optimal biomass"; AGB correlation of 0.7905. |
| Yuan et al. [86] | Textile engineering | Objective fabric smoothness assessment | 3D/2D $\alpha$-shape area ratio, multi-view stitching | Achieved 95.81% accuracy on 730 scans; $\alpha$-shape descriptor tracked expert evaluation; highlighted dataset, subjectivity, lab-condition, and hardware constraints. |
| Dey [87] | LiDAR buildings | Adaptive roof-boundary extraction | Adaptive $\alpha$-shape, scanline $\alpha$-selection, Gaussian smoothing, DP | Adaptive $\alpha$ improved relative area error under varying density; scanline-aware mechanism handled density variability. improved boundary extraction accuracy by 40% vs. CAA method. |
| Yan et al. [88] | Ocean engineering, | Automated detection of corrosion cavities | Adaptive $\alpha$-shape, outlier removal, cloth simulation filter | Adaptive $\alpha$-shape dynamically tuned to local point-cloud geometry; preserved concave and multi-scale corrosion-cavity shapes despite heavy sonar noise; avoided over-smoothing or fragmentation. |

Prior distance-based reconstruction methods typically compute the distance from each query point to an explicitly known physical boundary. Although this facilitates interior–exterior classification, it increases computational cost, especially for curved geometries where gradient-based normal estimation is often needed. As the first novelty, the proposed approach removes these dependencies by avoiding explicit boundary information and boundary-distance evaluations. Instead, it classifies structured-grid points via direct NN comparisons with the input samples, recasting the problem as a distance evaluation between two-point sets. A key limitation of the classical $\alpha$-shape method is its use of a single global $\alpha$, which often fails for complex boundaries where different regions require different $\alpha$ values, large for smooth convex parts and small for sharp or concave features. To address this, an adaptive $\alpha$-shape formulation is introduced as the second novelty that computes spatially varying $\alpha$ values directly from the Delaunay tessellation, using local mean edge-length statistics near the boundary. This data-driven adaptation improves agreement between the reconstructed boundary and the true physical geometry. The third novelty is a method-agnostic boundary inflation refinement that corrects discretization-driven boundary under-coverage on Cartesian grids. After reconstruction, the obtained mask is minimally dilated by an expansion factor $\eta$ to enclose boundary-adjacent samples that can be missed due to rasterization and sub-voxel alignment, thereby reducing edge misses while avoiding unnecessary activation of unsupported regions. Furthermore, a topology-aware set of metrics is proposed to quantify retention, unsupported-region suppression, overlap agreement, and connectivity. Beyond these methodological contributions, a dedicated web application is developed to operationalize the workflow from data import to CNN-ready output generation. Users can upload 2D ASCII datasets, tune the parameters of each reconstruction strategy, and automatically export structured outputs. The platform also identifies and reconstructs the physical boundary directly from the input data. This interactive interface enables straightforward application of distance-based classification, classical $\alpha$-shape reconstruction, and adaptive $\alpha$-shape formulation, supporting reproducible boundary recovery and efficient data preparation for machine-learning tasks.

Although the proposed methodology is extendable to higher-dimensional settings, the present evaluation is restricted to 2D geometries because the targeted CNN workflow operates on image-based inputs. This choice enables the reconstruction, masking, and metric-based assessment steps to be presented and validated in a direct and interpretable manner while remaining representative of many engineering cases. For 3D configurations, the same pipeline can be applied through a slice-based strategy: the 3D domain is decomposed into 2D cross-sections, each slice is processed with the identical reconstruction and post-processing operations, and the resulting stack can be used either as a set of 2D training samples or reassembled into a volumetric representation for further analysis. The remainder of the article is organized as follows. Section 2 introduces the problem statement and the evaluation configurations, and motivates domain reconstruction. Sections 3.1 through 3.5 present the full reconstruction pipeline, starting with grid construction and interpolation, followed by the classical baseline and the two proposed boundary recovery strategies, and concluding with the assessment metrics. Section 4 then reports and discusses the results: Section 4.1 analyzes how the classical $\alpha$-shape parameter affects boundary recovery and the inclusion of spurious regions, Sections 4.2 and 4.3 examine how the novel distance threshold and the adaptivity control parameter govern geometric fidelity, Section 4.4 compares the approaches in terms of mask generation runtimes, and Section 4.5 introduces boundary inflation as a post-processing step to mitigate residual edge misses. Finally, Section 5 summarizes the main findings. A brief tutorial for the integrated web application is provided in the Supplementary Material.

## 2. Problem Statement

Internal flows in confined geometries underpin many engineering and biomedical systems, including pipe networks, heat exchangers, nozzles, combustors, and vascular devices. Despite their apparent simplicity, they can exhibit separation, recirculation, secondary vortices, and strong acceleration and deceleration. These phenomena govern pressure losses, mixing, heat and mass transfer, and particle transport, and therefore demand accurate numerical prediction.

Figure 1a-d presents the four internal-flow configurations considered in this study. The first geometry, shown in Figure 1a, is a sudden expansion and contraction duct. A straight inlet section of length $L_i$ and width $R_i$ expands abruptly into a main section of length $L_m$ and width $R_m$, followed by a sudden contraction into an outlet section of length $L_o$ and width $R_o$. This expansion–contraction arrangement is a widely used canonical geometry for internal flows with abrupt area changes. The second configuration, illustrated in Figure 1b, is a Y-shaped bifurcating channel with a total branching angle of $2\phi$. The inlet branch has length $L_i$ and width $R_i$, and it splits at the junction into two daughter branches, each characterized by length $L_o$ and width $R_o$. This geometry represents flow division in engineering manifolds and idealized biological bifurcations. Figure 1c presents the third configuration which is a converging–diverging nozzle. The inlet converging section with length $L_i$ and width $R_i$ narrows toward a throat of minimum width $R_m$, followed by a diverging section with length $L_o$ and width $R_o$. This canonical nozzle layout is commonly used in propulsion, metering, and wind-tunnel test sections. Figure 1d depicts the fourth configuration, a curved turbine passage with upstream and downstream ducts. The upstream duct has length $L_i$ and width $R_i$, followed by a blade passage defined by the blade chord $L_c$, axial chord $L_{ac}$, inlet blade angle $\omega$, and exit blade angle $\varphi$, and finally a downstream duct of length $L_o$ and width $R_o$. This geometry captures the essential dimensional features of curved passages with blade-induced turning. Since the main objective of this paper is to accurately identify the physical boundaries of the problem and reconstruct the computational domain on a structured CNN grid, all geometric dimensions mentioned in Figure 1a-d have been adopted from previously validated studies to preserve their practical and engineering relevance, and are summarized in Table 3.

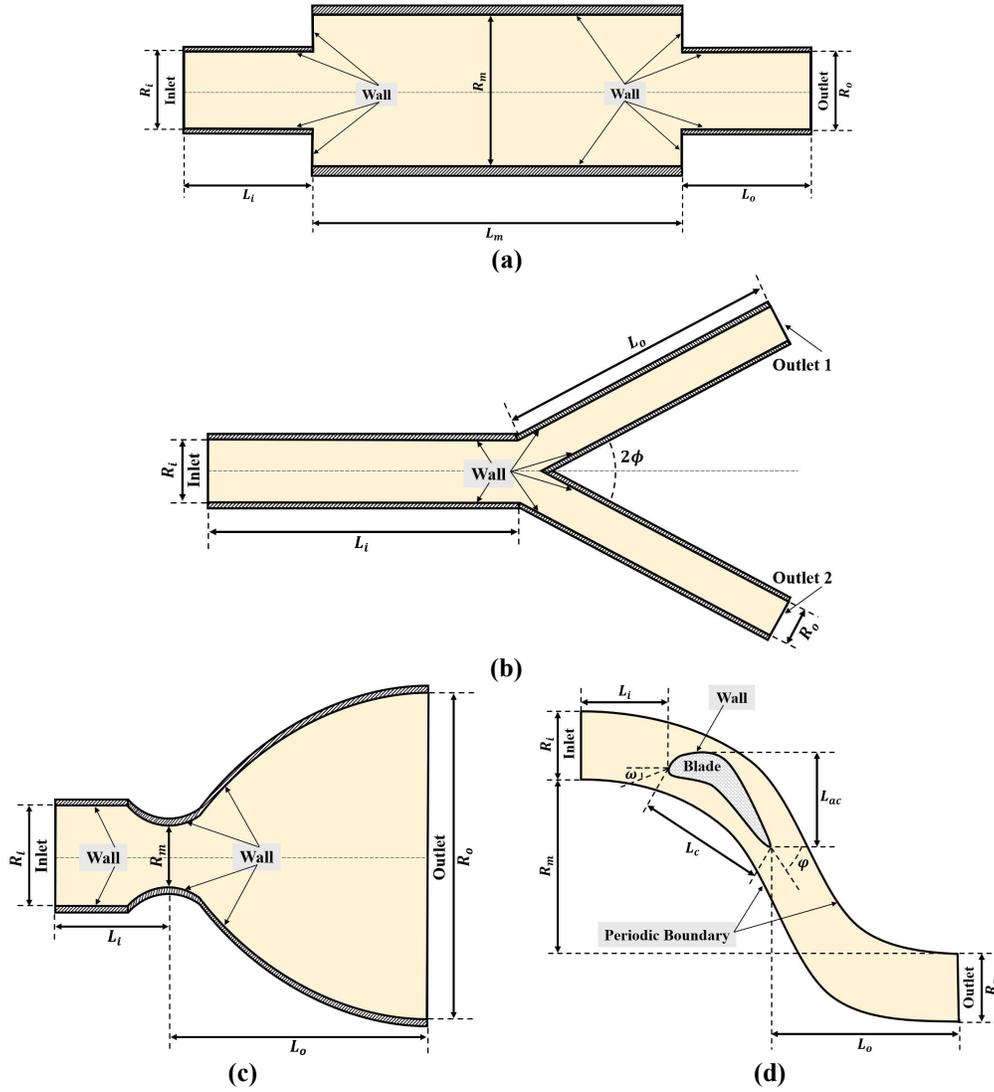

**Figure 1** Schematic of the internal flow configurations considered in this study: (a) Sudden expansion–contraction duct, (b) Y-shaped bifurcating channel, (c) Converging–diverging nozzle, and (d) Curved turbine passage.

**Table 3** Geometric dimensions of the internal flow configurations shown in Figure 1a-d.

| Configuration | $R_i(m)$ | $R_m(m)$ | $R_o(m)$ | $L_i(m)$ | $L_m(m)$ | $L_o(m)$ | $L_c(m)$ | $L_{ac}(m)$ | $2\phi(°)$ | $\omega(°)$ | $\varphi(°)$ |
|---|---|---|---|---|---|---|---|---|---|---|---|
| Sudden expansion–contraction duct [89] | 0.025 | 0.050 | 0.025 | 0.2 | 0.3 | 0.2 | – | – | – | – | – |
| Y-shaped bifurcating channel [90] | 0.016 | – | 0.012 | 0.058 | – | 0.050 | – | – | 49.4 | – | – |
| Converging–diverging nozzle [91] | 0.80 | 0.63 | 2.37 | 0.78 | – | 2.59 | – | – | – | – | – |
| Curved turbine passage [92, 93] | 0.030 | 0.026 | 0.030 | 0.022 | – | 0.043 | 0.047 | 0.037 | – | 48.8 | 79.5 |

After completing the CFD simulations for the selected geometries, the next step is to interpolate the solution fields, including the velocity components, turbulent kinetic energy, specific turbulence dissipation rate, temperature, and heat flux components, onto a uniform Cartesian grid so that the data can be used in the CNN framework. At this stage, however, a critical challenge arises, as illustrated in Figure 2a-d. Once any of these variables is mapped onto the uniform grid, the geometric representation is distorted, activating spurious, non-physical regions within the domain. Consequently, a robust reconstruction strategy is required to convert the convex envelope that naturally emerges from the grid interpolation process into a concave boundary that accurately recovers the actual physical domain.

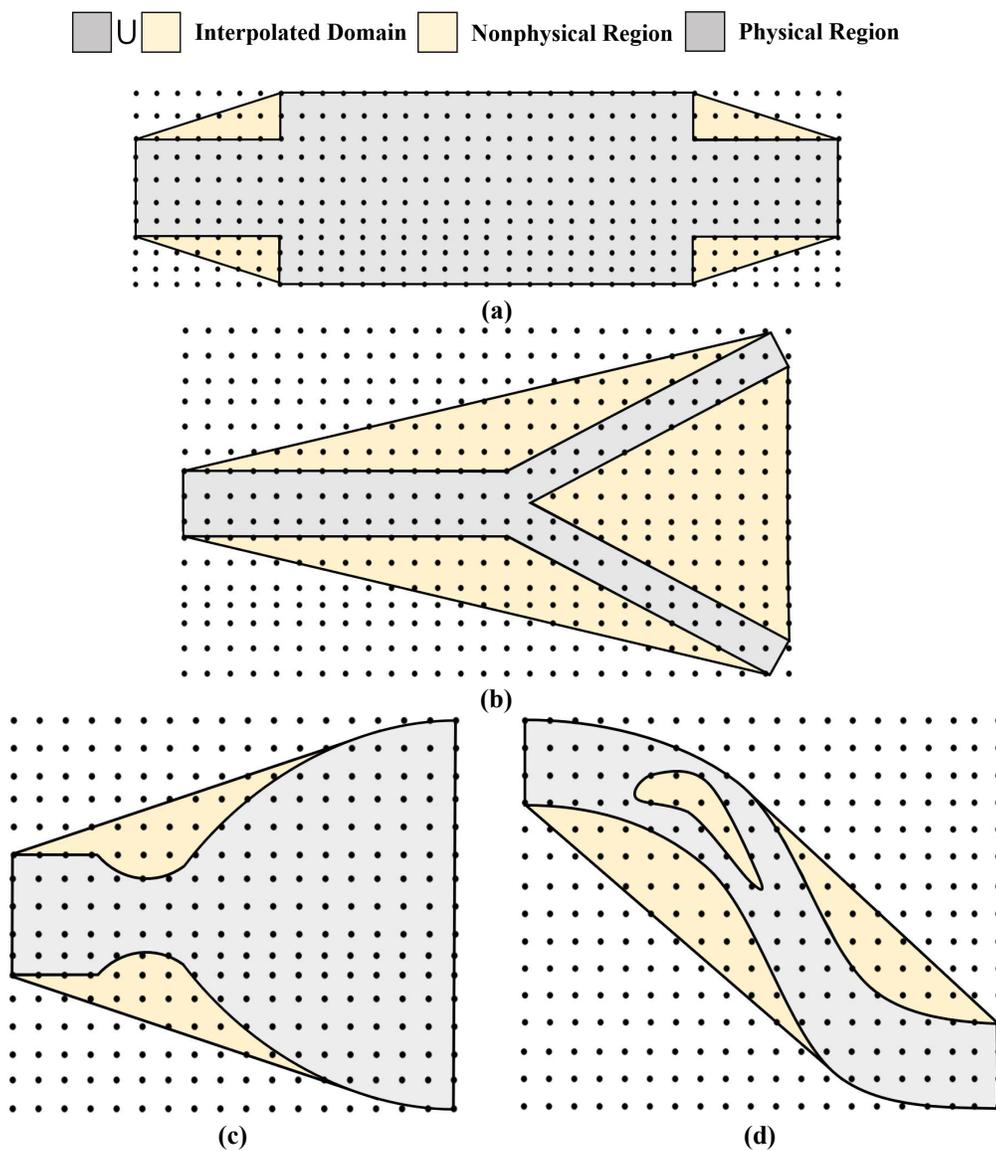

**Figure 2** Comparison of the interpolated CFD data on the uniform Cartesian grid, showing the physical domains in gray and the nonphysical regions in yellow.

## 3. Methodology

This section presents three distinct reconstruction strategies formulated to generate a structured representation of scattered data while accurately identifying the physical domain boundaries. After introducing the common preliminaries—namely, the definition of data points, the construction of a uniform Cartesian grid, and the piecewise-linear interpolation procedure—three distinct reconstruction methods are described: (i) the proposed distance-based reconstruction, (ii) the classical $\alpha$-shape reconstruction baseline, and (iii) a novel adaptive $\alpha$-shape formulation incorporating scaling with respect to local data resolution. Subsequently, a set of quantitative metrics is introduced to assess these reconstruction methods in terms of data preservation, geometric fidelity, and topological consistency on the structured grid.

### 3.1. Grid Construction and Interpolation

The set of scattered data samples, illustrated in Figure 3a, is defined as:

$$\mathcal{P} = \{x_i, f_i\}_{i=1}^{N}, \qquad x_i \in \mathbb{R}^d, \qquad f_i \in \mathbb{R}^q, \tag{1}$$

where $N$ represents the total number of samples, $d \in \{2,3\}$ is the spatial dimension, and $f_i$ collects either a scalar ($q = 1$) or multiple components ($q > 1$).

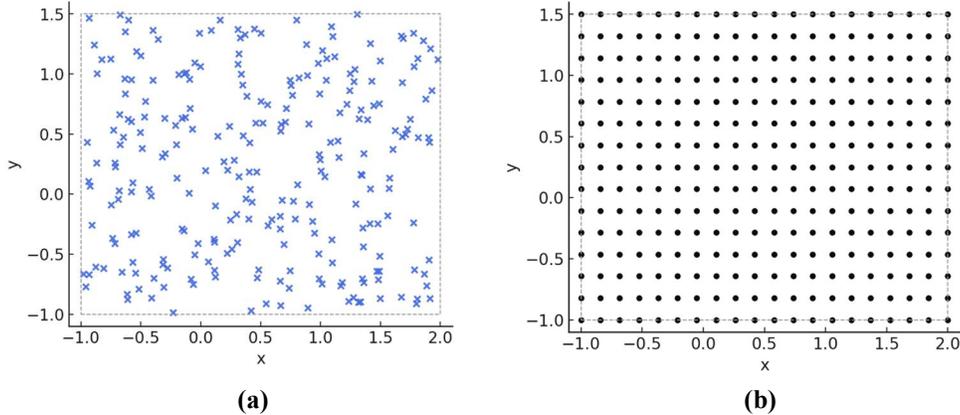

(a)          (b)

**Figure 3** Spatial distribution of (a) the scattered data samples and (b) the corresponding uniform grid representation.

An axis-aligned bounding box of the data is defined as:

$$\mathcal{B} = \{x \in \mathbb{R}^d | x_{min} \le x \le x_{max}\}, \tag{2}$$

with componentwise inequalities. As shown in Figure 3b, a structured grid covering $\mathcal{B}$ is constructed as the tensor product grid [94]:

$$\mathcal{G} = \{x'_j | j = (j_1, \ldots, j_d) \in \mathbb{N}^d, \mathbf{1} \leq j \leq n\}, \quad n = (n_1, \ldots, n_d), \tag{3}$$

where the coordinates of the grid points are uniformly spaced as [94]:

$$x'_j = x_{min} + (j - \mathbf{1}) \odot \Delta x, \quad \Delta x = \frac{x_{max} - x_{min}}{n - \mathbf{1}}. \tag{4}$$

Here, $\odot$ denotes the Hadamard product, $n$ is the CNN grid-resolution vector, $\mathbf{1}$ is the all-ones vector, and the division in $\Delta x$ is interpreted componentwise.

The scattered data are interpolated onto the structured grid using a piecewise-linear interpolant of the form as [95]:

$$f^*(x') = \sum_{i=1}^{N} \gamma_i(x') f_i, \quad \sum_{i=1}^{N} \gamma_i(x') = 1. \tag{5}$$

The weights $\gamma_i(x')$ are defined with respect to the Delaunay tessellation of $\{x_i\}_{i=1}^{N}$ [96-98] and are nonzero only for the vertices of the unique simplex that contains $x'$. Hence, for any query point $p' \in \mathbb{R}^d$, at most $d + 1$ samples contribute to $f^*(x')$ [99, 100], as illustrated in Figure 4.

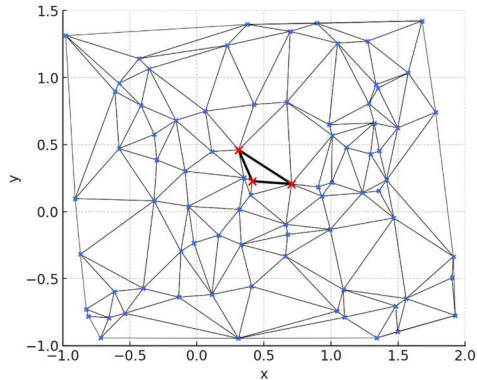

**Figure 4** Delaunay triangulation of the scattered data samples, used to define local simplices for piecewise-linear interpolation.

In particular, if $p'$ lies in the Delaunay simplex $\sigma \coloneqq \text{conv}(x_{i_0}, \ldots, x_{i_d})$, the interpolant reduces to the following barycentric form [101, 102]:

$$f^*(p') = \sum_{m=0}^{d} \lambda_m(p') f_{i_m}, \tag{6}$$

where $\lambda_m(\boldsymbol{p}')$ denotes the barycentric coordinate of $\boldsymbol{p}'$ with respect to $\sigma$, satisfying $\sum_{m=0}^{d}\lambda_m(\boldsymbol{p}') = 1$ and $\lambda_m(\boldsymbol{p}') \geq 0$ when $\boldsymbol{p}' \in \sigma$ [103].

To evaluate Eq. (6), the barycentric coordinates of $\boldsymbol{p}'$ in its enclosing simplex are computed from the vertex coordinates, as illustrated in Figure 5. An affine representation is given by:

$$\boldsymbol{p}' = \sum_{m=0}^{d} \lambda_m \boldsymbol{x}_{i_m}, \qquad \sum_{m=0}^{d} \lambda_m = 1, \qquad \boldsymbol{x}_{i_m} \in \mathbb{R}^d. \tag{7}$$

The weight vector $\boldsymbol{\lambda} = (\lambda_0, \dots, \lambda_d)^T$ can be obtained by solving the linear system [101, 104]:

$$\begin{bmatrix} \boldsymbol{x}_{i_0} & \boldsymbol{x}_{i_1} & \cdots & \boldsymbol{x}_{i_d} \\ 1 & 1 & \cdots & 1 \end{bmatrix} \boldsymbol{\lambda} = \begin{bmatrix} \boldsymbol{p}' \\ 1 \end{bmatrix}. \tag{8}$$

An alternative but equivalent form defines the barycentric coordinates using simplex measure ratios as [101, 104]:

$$\lambda_m(\boldsymbol{p}') = \frac{|\sigma_m(\boldsymbol{p}')|_{(a-v)}}{|\sigma|_{(a-v)}}, \qquad \sigma_m(\boldsymbol{p}') := \mathrm{conv}(\boldsymbol{x}_{i_0}, \dots, \boldsymbol{x}_{i_{m-1}}, \boldsymbol{p}', \boldsymbol{x}_{i_{m+1}}, \dots, \boldsymbol{x}_{i_d}). \tag{9}$$

Here $|\cdot|_{(a-v)}$ represents area in 2D and volume in 3D. The simplex measure may be computed compactly as follows [101, 104]:

$$|\sigma|_{(a-v)} = \frac{1}{d!} |\det[\boldsymbol{x}_{i_1} - \boldsymbol{x}_{i_0}, \dots, \boldsymbol{x}_{i_d} - \boldsymbol{x}_{i_0}]|, \tag{10}$$

The interpolated field $\boldsymbol{f}^*(\boldsymbol{p}')$ defined by Eq. (6) provides a smooth and continuous representation of the scattered dataset, forming the basis for the subsequent reconstruction procedures [105].

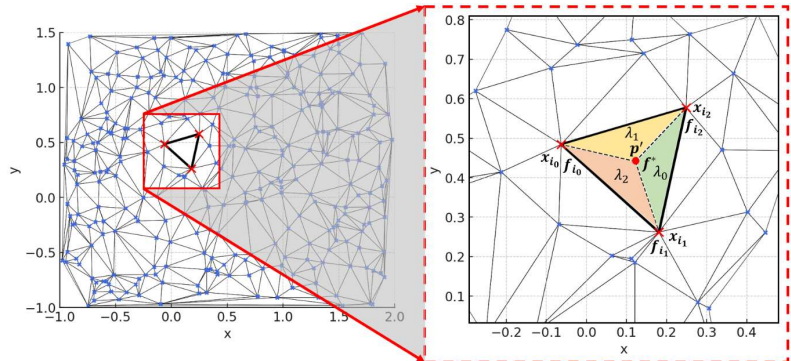

**Figure 5** Geometric interpretation of barycentric interpolation within the enclosing Delaunay simplex.

*3.2. Distance-Based Reconstruction*

The primary objective of this and the subsequent sections is to refine the convex-hull envelope, which naturally results from grid interpolation, into a concave boundary that accurately recovers the true physical geometry. In this section, this is achieved through the proposed distance-based masking strategy, which compares the uniform Cartesian grid nodes against the original scattered samples to identify and suppress the nonphysical regions introduced during interpolation, as illustrated in Figure 6.

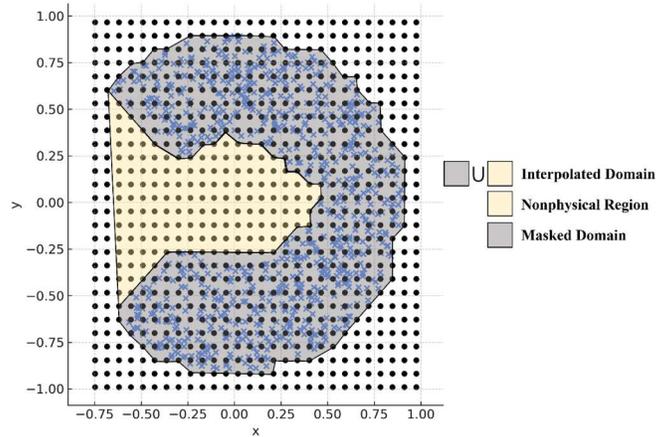

**Figure 6** Comparison between the scattered samples and the uniform grid after masking, highlighting the physical domain (gray) and the nonphysical region (yellow).

For each grid node $x'_j \in \mathcal{G}$, the Euclidean distance to the nearest sample point is computed as [106]:

$$D(x'_j) = \min_{1 \leq i \leq N} \left\| x'_j - x_i \right\|_2, \tag{11}$$

where $\|\cdot\|_2$ denotes the Euclidean norm in $\mathbb{R}^d$. Collecting these values over all grid nodes yields the distance map $D = \{D(x'_j)\}$, which remains small in regions densely supported by samples and grows in sparsely sampled or potentially nonphysical regions [54, 107], as shown in Figure 7.

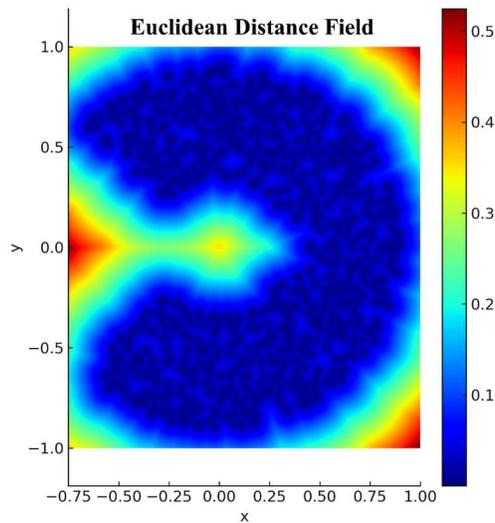

**Figure 7** Euclidean distance field showing smaller values (blue) in well-sampled regions and larger values (red) near nonphysical areas.

A binary indicator function is then defined to identify grid nodes that lie within a user-specified threshold $\tau > 0$:

$$\chi_{db}(x'_j) = \begin{cases} 1, & \text{if } D(x'_j) < \tau, \\ 0, & \text{otherwise,} \end{cases} \tag{12}$$

which selects a conservative tubular neighborhood around the sample cloud. Choosing $\tau$ too small may yield under-coverage and internal voids in sparsely sampled regions, whereas overly large $\tau$ can extend beyond outside the true boundary, particularly near concavities. A practical default is $\tau = \min_{1 \leq i \leq d}(\Delta x_i)$, which ensures consistency with the grid resolution and provides an appropriate balance between boundary accuracy and geometric completeness [51, 108].

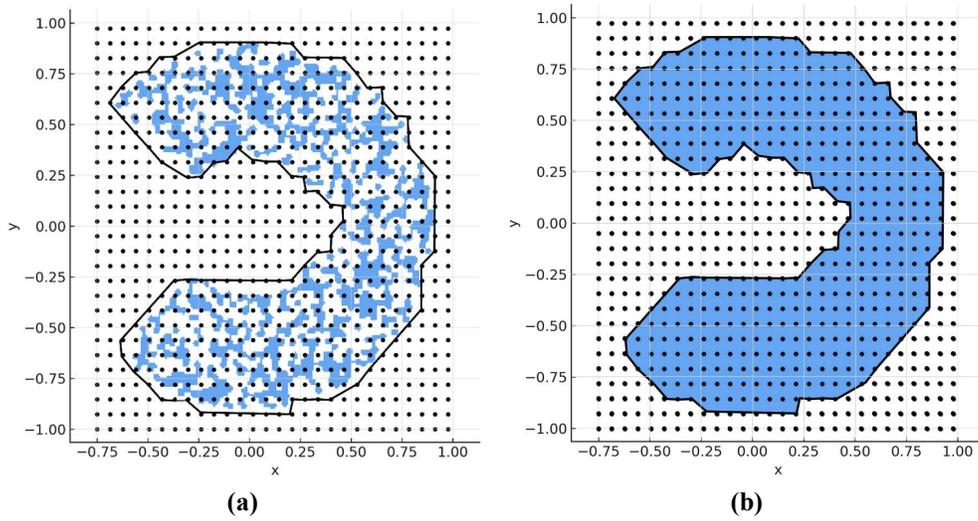

(a)  (b)
**Figure 8** Morphological closing of the distance-based mask: (a) Preliminary in-domain voxels $M_{db}$ and (b) Refined voxel set $M_{db,f}$.

To remove any nonphysical holes or disconnected regions in the preliminary mask $M_{db} = \{x'_j \in \mathcal{G} | \chi_{db}(x'_j) = 1\}$, as shown in Figure 8a, and to obtain a smoother, contiguous representation, a morphological closing is applied to $\chi_{db}$ defined in Eq. (12) as [51, 109]:

$$\chi_{db,f} = (\chi_{db} \oplus B) \ominus B, \tag{13}$$

where $\oplus$ and $\ominus$ denote binary dilation and erosion, respectively. The dilation first expands $\chi_{db}$ outward by up to $r_c = \frac{s-1}{2}$ voxels in the Chebyshev sense, closing narrow gaps and seams; the subsequent erosion restores the outer boundary while preserving the newly bridged connections and filled pinholes [110, 111].

The structuring element $B$ is taken as a cubic neighborhood [110, 112]:

$$B = \{(a, b, c) \in \mathbb{Z}^3 | |a| \leq r_c, |b| \leq r_c, |c| \leq r_c\}, \tag{14}$$

where $(a, b, c)$ are dimensionless integer voxel offsets in index space. The parameter $r_c = \frac{s-1}{2}$ is the Chebyshev ($L_\infty$) radius, equal to half the cube side length, and $s$ is an odd integer with $s \geq 1$ [113]. Larger $s$ values close wider gaps but may over-smooth thin features; smaller values better preserve detail but may leave isolated holes. As depicted in Figure 8b, the resulting set of in-domain voxels, $M_{db,f} = \{x'_j \in \mathcal{G} | \chi_{db,f}(x'_j) = 1\}$, conforms more closely to the physical geometry than $M_{db}$ and is more robust to local sampling irregularities [114, 115].

### 3.3. $\alpha$-Shape Reconstruction

Following the distance-based approach, the second method, referred to as the classical $\alpha$-shape reconstruction, provides a mathematically rigorous framework for recovering the true concave boundaries of the physical domain [116, 117]. Unlike distance thresholding, which operates locally in the Euclidean sense, the $\alpha$-shape approach globally reconstructs the domain boundary based on topological and geometric relations between data points. It effectively generalizes the convex hull by introducing a tunable length-scale parameter $\alpha$ that controls the degree of boundary concavity [118].

Given the data point set $\mathcal{P}$ in Eq. (1), a Delaunay complex $\text{Del}(\mathcal{P})$ is first constructed to establish the connectivity among neighboring samples [99, 119], as illustrated in Figure 4. Each simplex $\sigma := \text{conv}(x_{i_0}, \dots, x_{i_d})$ in $\text{Del}(\mathcal{P})$ is associated with a unique empty circumball $S(c_\sigma, r_sigma)$ that satisfies the following relations [120]:

$$\|c_\sigma - x_{i_m}\|_2 = r_\sigma, \quad m = 0, \dots, d, \tag{15}$$

$$\|c_\sigma - x\|_2 > r_\sigma, \quad \forall x \in \mathcal{P} \setminus \{x_{i_0}, \dots, x_{i_d}\}. \tag{16}$$

Here, $c_\sigma$ and $r_\sigma$ correspond to the circumcenter and circumradius of $\sigma$, respectively.

Given a specified scale parameter $\alpha > 0$, the $\alpha$-complex $C_\alpha \subseteq \text{Del}(\mathcal{P})$ is defined as the subcomplex containing all Delaunay simplices whose circumradius does not exceed $\alpha$ as follows [121], as illustrated in Figure 9:

$$C_\alpha = \{\sigma \in \text{Del}(\mathcal{P}) | r_\sigma \leq \alpha\}. \tag{17}$$

To ensure topological closure, all faces of each selected simplex are also included. In practice, $\alpha$ can be linked to the spatial resolution of the structured grid, which is given by [122]:

$$\alpha \approx \delta \left[ \min_{1 \leq i \leq d} (\Delta x_i) \right], \quad 0.5 \leq \delta \leq 2, \tag{18}$$

where $\delta$ is an empirical scaling coefficient. Smaller $\alpha$ values preserve fine boundary undulations but can fragment thin or sparsely sampled regions, whereas larger values produce smoother reconstructions at the expense of local geometric detail [123].

The $\alpha$-shape of the point set $\text{Sh}_\alpha(\mathcal{P})$ is then defined as the geometric realization of this complex as [68]:

$$\text{Sh}_\alpha(\mathcal{P}) = \bigcup_{\sigma \in C_\alpha} \sigma. \tag{19}$$

Its boundary $\partial \text{Sh}_\alpha(\mathcal{P})$ provides the reconstructed geometry of the physical domain, as demonstrated in Figure 9. As $\alpha$ varies, the resulting shape continuously transitions between the discrete data cloud and its convex hull as follows [123]:

$$\lim_{\alpha \to 0^+} \text{Sh}_\alpha(\mathcal{P}) = \mathcal{P}, \tag{20}$$

$$\lim_{\alpha \to \infty} \text{Sh}_\alpha(\mathcal{P}) = \text{conv}(\mathcal{P}). \tag{21}$$

Moreover, the $\alpha$-complex and $\alpha$-shape are topologically consistent through the relationship below:

$$\partial C_\alpha(\mathcal{P}) = \partial \text{Sh}_\alpha(\mathcal{P}), \tag{22}$$

which ensures that $\text{Sh}_\alpha(\mathcal{P})$ is a bounded polytope whose boundary is formed of the exposed simplices of $C_\alpha(\mathcal{P})$ [68].

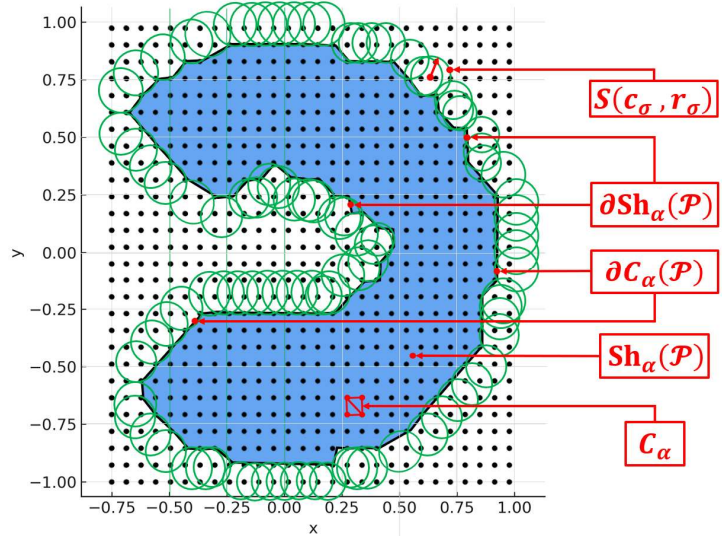

**Figure 9** $\alpha$-shape reconstruction depicting the $\alpha$-complex $C_\alpha$, circumcircles $S(x_\sigma, r_\sigma)$, and resulting $M_{\alpha,f}$ defined by $\text{Sh}_\alpha(\mathcal{P})$ and its boundary $\partial \text{Sh}_\alpha(\mathcal{P})$.

For computational implementation, $r_\sigma$ used in Eqs. (15) to (17) can be expressed analytically. For a Delaunay triangle with vertices $x_{i_0}$, $x_{i_1}$, $x_{i_2}$ and side lengths $e_{2,1} = \|x_{i_1} - x_{i_2}\|_2$, $e_{2,0} = \|x_{i_0} - x_{i_2}\|_2$, and $e_{1,0} = \|x_{i_0} - x_{i_1}\|_2$, the circumradius is given by [124, 125]:

$$r_\sigma = \frac{e_{2,1} e_{2,0} e_{1,0}}{4 A_{0,1,2}}, \tag{23}$$

where $A_{0,1,2}$ is the triangle area, computed using the $|\cdot|_{(a-v)}$ operator defined in Eq. (10).

Once $\text{Sh}_\alpha(\mathcal{P})$ is obtained, its polygonal boundary defines a geometric mask that confines the interpolation grid. Each grid node defined in Eqs. (2) to (4) is then classified as:

$$\chi_\alpha(x_j') = \begin{cases} 1, & \text{if } x_j' \in \text{Sh}_\alpha(\mathcal{P}), \\ 0, & \text{otherwise}, \end{cases} \tag{24}$$

yielding the final in-domain set $M_{\alpha,f} = \{x_j' \in \mathcal{G} | \chi_\alpha(x_j') = 1\}$, as shown in Figure 9. This approach naturally accommodates both external concavities and internal cavities, eliminating the need for heuristic thresholds or morphological filters [126, 127].

### 3.4. Adaptive α-Shape Reconstruction

Although the classical $\alpha$-shape reconstruction provides a reliable topological framework, its performance depends strongly on the manually prescribed scale parameter $\alpha$. Since $\alpha$ is specified as an absolute length scale in the coordinate units of the dataset, changes in sampling density or coordinate scaling can lead to inconsistent boundary reconstructions [128, 129]. To mitigate this limitation, a novel adaptive and resolution-independent formulation is introduced in this section.

Building upon the set of data points defined in Eq. (1) and the corresponding Delaunay tessellation [99, 119] illustrated in Figure 4, the set of all unique edges is expressed as:

$$\mathcal{E} = \{e_1, \ldots, e_{N_e}\}, \tag{25}$$

where $N_e$ is the total number of unique edges. A characteristic length scale is then defined as the mean Delaunay edge length [117]:

$$\bar{e} = \max\left(\frac{1}{N_e} \sum_{e=1}^{N_e} \|x_{e,1} - x_{e,2}\|_2, 10^{-12}\right), \tag{26}$$

with $x_{e,1}$ and $x_{e,2}$ denoting endpoints of edge $e$. The adaptive $\alpha$ parameter, $\alpha_a$, is subsequently normalized by this length scale as [130]:

$$\alpha_a = \frac{\beta}{\bar{e}}, \tag{27}$$

where $\beta > 0$ is a dimensionless control parameter that regulates boundary concavity independently of absolute geometric units.

For each simplex $\sigma \in \text{Del}(\mathcal{P})$, characterized by its circumball $S(c_\sigma, r_\sigma)$ satisfying Eqs. (15) and (16), the subset of simplices retained in the adaptive $\alpha$-complex is identified as [131]:

$$C_{a\alpha} = \{\sigma \in \text{Del}(\mathcal{P}) | r_\sigma \leq \alpha_a^{-1}\}, \tag{28}$$

so that only well-shaped simplices with small circumradii are preserved. Each retained simplex $\sigma \in C_{a\alpha}$ contains a set of boundary elements, denoted by $EF_\sigma$ [67]. The occurrence count of each boundary element, $ef$, across all retained simplices is computed as:

$$n_{ef} = \sum_{\sigma \in C_{a\alpha}} \mathbf{1}(ef \in EF_\sigma), \tag{29}$$

where $\mathbf{1}(\cdot) \in \{0,1\}$ is the counting function [132]. Boundary elements appearing only once define the external boundary as expressed below [131]:

$$EF_{eb} = \{ef \in \cup_\sigma EF_\sigma \,|\, n_{ef} = 1\}. \tag{30}$$

The adaptive $\alpha$-shape is obtained by taking the union of the convex hulls of these exposed boundary elements:

$$\text{Sh}_{a\alpha}(\mathcal{P}) = \bigcup_{ef \in EF_{eb}} \text{conv}(ef), \tag{31}$$

which forms a closed polygonal curve, representing the boundary $\partial \text{Sh}_{a\alpha}(\mathcal{P})$ of the reconstructed domain [131].

At the end, structured grid nodes are classified using the same rule as in Eq. (24), resulting in the final in-domain set $M_{a\alpha,f} = \{x'_j \in \mathcal{G} | \chi_{a\alpha}(x'_j) = 1\}$, which delineates the physically valid portion of the structured grid. This adaptive formulation preserves the theoretical rigor of the $\alpha$-shape framework while enhancing its robustness, reproducibility, and geometric consistency across datasets with varying scales and sampling densities [117, 128, 129]. It therefore provides a reliable and resolution-independent tool for domain reconstruction suitable for structured-grid representation and CNN-based analysis.

### 3.5. Reconstruction Assessment Metrics

To objectively evaluate the proposed reconstruction strategies against the classical baseline, a set of quantitative metrics is introduced in this section. Taken together, these criteria assess reconstruction fidelity and topology by measuring how well each method retains the original scattered samples, suppresses unsupported or extrapolated regions, and reproduces the expected geometry and connectivity of the physical domain on the structured grid. Table 4 lists the complete set of metrics evaluated for each method $k \in \{db, \alpha, a\alpha\}$, where $db$, $\alpha$, and $a\alpha$ refer to the distance-based, classical $\alpha$-shape, and adaptive $\alpha$-shape methods, respectively.

**Table 4** Quantitative criteria for evaluating reconstruction accuracy and topology.

| Number | Criterion | Symbol | Definition | What it measures |
|---|---|---|---|---|
| 1 | Point Recall | $PR_k$ | $\frac{1}{N}\sum_{i=1}^{N}\chi_k(\pi(x_i))$ | Boundary data retention |
| 2 | Ghost Fraction | $GF_k$ | $\frac{|\mathcal{G}_k|}{|M_k|} = \frac{\sum_{j\in\mathcal{L}}\mathbf{1}[\chi_k(j)=1,\ D(x'_j)>r_0]}{\sum_{j\in\mathcal{L}}\mathbf{1}[\chi_k(j)=1]}$ | Unsupported active region |
| 3 | Active Volume Fraction | $AVF_k$ | $\frac{N_{av,k}}{|\mathcal{L}|} = \frac{\sum_{j\in\mathcal{L}}\chi_k(j)}{\prod_{i=1}^{d}n_i}$ | Relative mask size |
| 4 | Connected Components | $N_{c,k}$ | $M_k = \bigcup_{i=1}^{N_{c,k}} CC_k^i$, $CC_k^i \cap CC_k^s = \emptyset\ (i \neq s)$ | Domain fragmentation |
| 5 | Intersection over Union | $IoU_k$ | $\frac{|\mathcal{I}_k|}{|\mathcal{U}_k|} = \frac{\sum_{j\in\mathcal{L}}\mathbf{1}[\chi_k(j)=1 \wedge \chi_\alpha(j)=1]}{\sum_{j\in\mathcal{L}}\mathbf{1}[\chi_k(j)=1 \vee \chi_\alpha(j)=1]}$ | Global overlap with classical $\alpha$-shape |
| 6 | Precision | $Pre_k$ | $\frac{|\mathcal{I}_k|}{|M_k|} = \frac{\sum_{j\in\mathcal{L}}\mathbf{1}[\chi_k(j)=1 \wedge \chi_\alpha(j)=1]}{\sum_{j\in\mathcal{L}}\mathbf{1}[\chi_k(j)=1]}$ | Leakage outside classical $\alpha$-shape |
| 7 | Recall | $Rec_k$ | $\frac{|\mathcal{I}_k|}{|M_\alpha|} = \frac{\sum_{j\in\mathcal{L}}\mathbf{1}[\chi_k(j)=1 \wedge \chi_\alpha(j)=1]}{\sum_{j\in\mathcal{L}}\mathbf{1}[\chi_\alpha(j)=1]}$ | Missing classical $\alpha$-shape coverage |

**Remarks:**
1. $PR_k$ measures the fraction of original samples mapped inside the reconstructed mask.
2. $GF_k$ quantifies active grid nodes farther than the reference radius $r_0$ from the data cloud, indicating unsupported or extrapolated, potentially nonphysical regions.
3. $AVF_k$ is the normalized active mask size relative to the full grid.
4. $N_{c,k}$ counts disconnected components, reflecting topological fragmentation.
5. $IoU_k$, $Pre_k$, and $Rec_k$ compare method $k$ against the classical $\alpha$-shape reference mask, where $IoU_k$ summarizes overall agreement, $Pre_k$ penalizes false positives, and $Rec_k$ penalizes false negatives.

In Table 4, $\chi_k(j) \in \{0,1\}$ denotes the binary indicator function associated with method $k$, which defines the active index set $M_k = \{j \in \mathcal{L}: \chi_k(j) = 1\}$. Here, $\mathcal{L} = \{1, \ldots, n_1\} \times \ldots \times \{1, \ldots, n_d\}$ is the structured grid index set, and $\pi(x_i)$ maps each sample location $x_i$ to its associated grid index. For overlap-based metrics ($IoU_k$, $Pre_k$, and $Rec_k$), the $\alpha$-shape mask is taken as the reference. Accordingly, the intersection and union sets are defined as $\mathcal{I}_k = M_k \cap M_\alpha$ and $\mathcal{U}_k = M_k \cup M_\alpha$, respectively. For the ghost fraction ($GF_k$), the reference radius is $r_0 = \rho\overline{d^{NN}}$, where $\rho > 0$ is a Neighborhood scaling factor and $\overline{d^{NN}}$ is the mean NN distance within the data cloud, computed as:

$$\overline{d^{NN}} = \frac{1}{N}\sum_{i=1}^{N} d_i^{NN} = \frac{1}{N}\sum_{i=1}^{N}\left(\min_{\substack{1\leq m\leq N \\ m\neq i}} \|x_i - x_m\|_2\right). \tag{32}$$

## 4. Results and Discussion

In this section, the performance and accuracy of each reconstruction method are carefully analyzed over a wide range of key parameters: $0.01 \leq \alpha \leq 7000$, $\frac{1}{8}\Delta \leq \tau \leq 8\Delta$, and $0.01 \leq \beta \leq 100$. For this purpose, a moderate-resolution CFD mesh with approximately 24000 to 36000 cells is used for all cases, ensuring that the discretization is representative of standard engineering simulations while remaining computationally manageable. Nevertheless, the proposed methods are general and can be applied to any arbitrary CFD mesh. Regarding the selection of the CNN grid, the optimal values of the model parameters are highly sensitive to grid resolution. Therefore, it is preferable to determine these values using a fine-resolution CNN grid so that the best parameter settings can be identified and subsequently applied to coarser CNN grids if needed. Accordingly, a fine grid of $n_x = n_y = 1000$ is adopted in this study.

### 4.1. Classical α-Shape Reconstruction: Effects of α

Across the evaluation configurations in Section 2, $\alpha$ parameter sets how strictly simplices are filtered by circumradius, and the metrics follow a consistent $\alpha$-dependence. At small $\alpha$, the reconstruction is permissive and can bridge concavities or gaps, yielding very high $PR_\alpha$ and large $AVF_\alpha$, but also elevated $GF_\alpha$ due to weakly supported voxels. As $\alpha$ increases to a moderate range, large circumradius simplices are removed and the boundary retracts toward data supported regions, so $GF_\alpha$ drops quickly and $AVF_\alpha$ contracts while $PR_\alpha$ typically stays near unity; topology is usually preserved and $N_{c,\alpha} \approx 1$. If $\alpha$ becomes too large, filtering is overly restrictive in thin or sparsely sampled regions, causing loss of connections and fragmentation, seen as sharp decreases in $PR_\alpha$ and $AVF_\alpha$ and a rise in $N_{c,\alpha}$. Because the metrics exhibit the same tradeoff structure with $\alpha$ across geometries, the detailed results for the first three cases (Figure 1a-c) are provided in Appendix. Therefore, as shown in Figure 10a and Figure 11a, only the most complex geometry, i.e. the **curved turbine passage**, is discussed here.

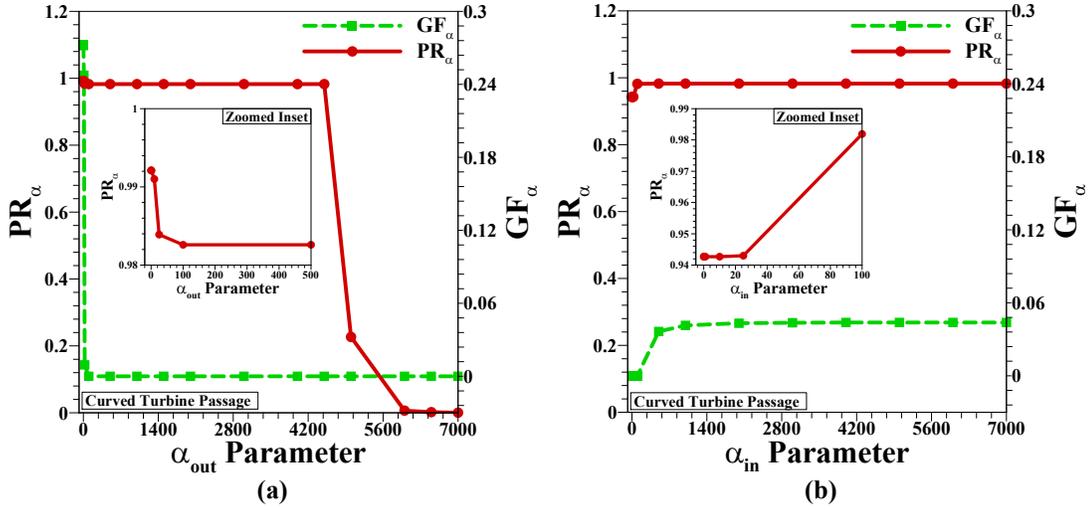

**Figure 10** Effect of the $\alpha$ parameter on $PR_\alpha$ and $GF_\alpha$ in the curved turbine passage: (a) Outer boundary and (b) Inner boundary.

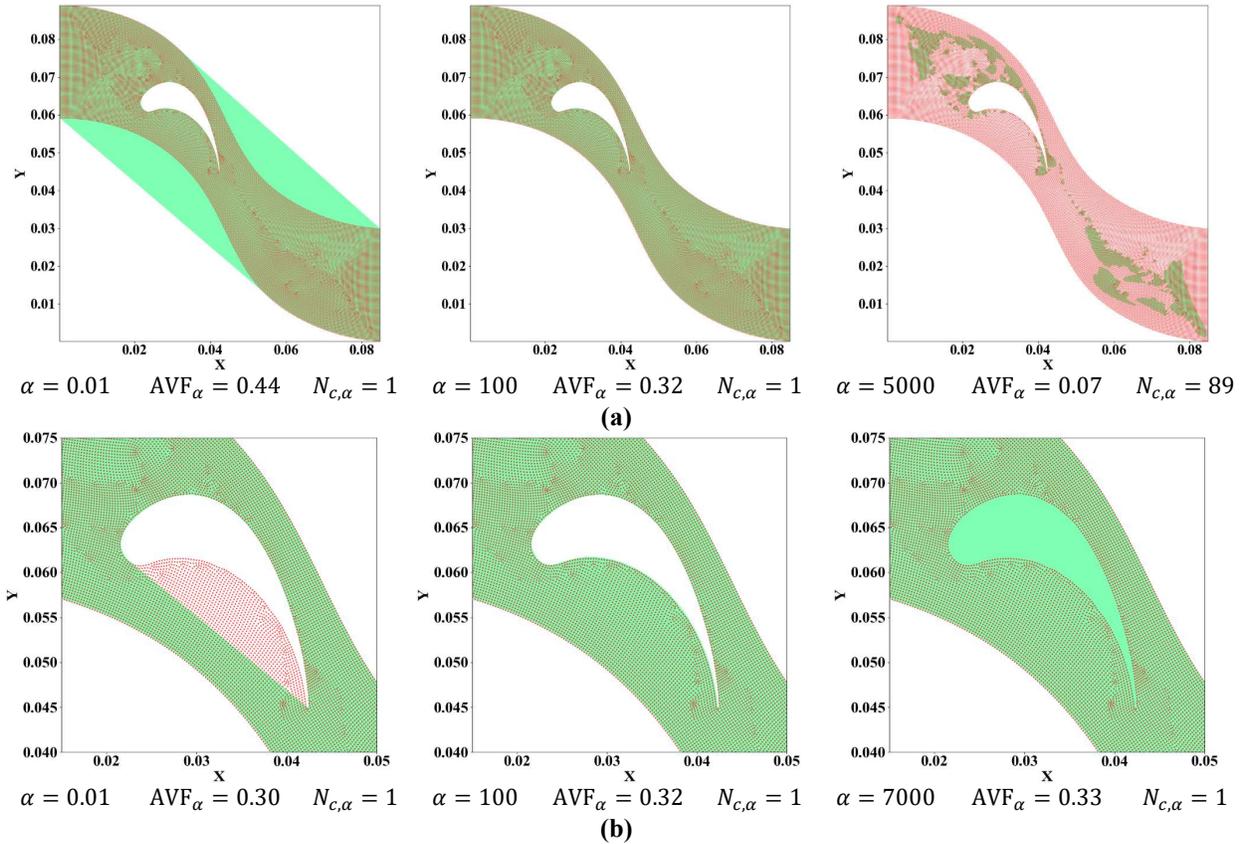

**Figure 11** $\alpha$-shape reconstruction outcome for the curved turbine passage: (a) Outer boundary and (b) Inner boundary. The green region denotes the reconstructed domain, and the red points represent the CFD data.

For the **curved turbine passage** outer boundary (Figure 10a and Figure 11a), $PR_\alpha$ remains high and nearly constant for $\alpha_{out} \leq 10$, then stays slightly lower but stable for $25 \leq \alpha_{out} \leq 4500$, before collapsing beyond $\alpha_{out} = 5000$. Over this range, $GF_\alpha$ decreases from about 0.272 at $\alpha_{out} = 0.01$ to 0.248 at $\alpha_{out} = 10$, drops to approximately 0 by $\alpha_{out} = 25$, and remains 0 for $\alpha_{out} \geq 100$. In parallel, $AVF_\alpha$ contracts from about 0.44 at very small $\alpha_{out}$ to about 0.32 by $\alpha_{out} = 100$ and then plateaus until the final collapse. The reconstruction remains topologically connected with $N_{c,\alpha} = 1$ up to $\alpha_{out} = 4500$, whereas $\alpha_{out} = 5000$ triggers severe fragmentation, evidenced by the sharp rise in $N_{c,\alpha}$ and the simultaneous loss in $PR_\alpha$ and $AVF_\alpha$. This identifies a broad, well-conditioned $\alpha_{out}$ window in which unsupported exterior activation is suppressed without compromising boundary connectivity. The internal boundary reconstruction for the **curved turbine passage** (Figure 10b and Figure 11b) exhibits a distinct sensitivity because the inner $\alpha$-shape defines an exclusion region that is subtracted from the grid mask. $PR_\alpha$ remains nearly constant at about 0.943 for $\alpha_{in} \leq 25$, then increases sharply to approximately 0.982 at $\alpha_{in} = 100$ and stays stable thereafter. $GF_\alpha = 0$ up to $\alpha_{in} = 100$, but rises to about 0.036 to 0.044 for $\alpha_{in} \geq 500$. Consistently, $AVF_\alpha$ is almost constant near 0.302 for $\alpha_{in} \leq 25$, increases to roughly 0.317 at $\alpha_{in} = 100$, and approaches about 0.333 at large $\alpha_{in}$. The cavity remains a single connected component for all $\alpha_{in}$, with $N_{c,\alpha} = 1$. Geometrically, small $\alpha_{in}$ yields an oversized inner complex that removes a larger-than-intended region, suppressing $PR_\alpha$ while keeping $GF_\alpha = 0$ because the operation is subtractive. Near $\alpha_{in} \approx 100$, the inner boundary aligns more closely with the sampled contour, recovering previously excluded samples and increasing $PR_\alpha$ and $AVF_\alpha$. At very large $\alpha_{in}$, the inner reconstruction becomes under-restrictive with respect to the intended cavity removal, leaving residual interior voxels that are weakly supported by nearby samples, which explains the increase in $GF_\alpha$ without a change in $N_{c,\alpha}$.

Table 5 Optimal $\alpha$ values identified from reconstruction metrics.

| Configuration | $\alpha_{out,opt}$ | $\alpha_{in,opt}$ |
|---|---|---|
| Sudden expansion–contraction duct | 100 | – |
| Y-shaped bifurcating channel | 1000 | – |
| Converging–diverging nozzle | 10 | – |
| Curved turbine passage | 100 | 150 |

The optimal $\alpha$ value for each geometry is reported in Table 5, and these values are subsequently employed in the next section to assess two further reconstruction methods.

*4.2. Distance-Based Reconstruction: Effects of $\tau$*

The distance-based reconstruction is controlled by the threshold $\tau$ in Eq. (12), which defines a buffer band of radius $\tau$ around the scattered point set when evaluated on the structured grid. In normalized form, $\frac{\tau}{\Delta}$ sets the effective mask thickness. For very small $\frac{\tau}{\Delta}$, the buffer band collapses to a narrow layer, so $PR_{db}$ is low because many boundary-adjacent samples fall outside the neighborhood. In this conservative regime, $GF_{db}$ is near 0 and $Pre_{db}$ is close to 1, indicating a strictly data-supported mask that lies largely within the reference $\alpha$-shape. As $\frac{\tau}{\Delta}$ increases from 0.25 to 0.5, $PR_{db}$ rises sharply toward 1 while $GF_{db}$ remains negligible, and $IoU_{db}$ and $Rec_{db}$ peak in a narrow range around $\frac{\tau}{\Delta} \approx 0.5$ to 1. For larger $\frac{\tau}{\Delta}$, $PR_{db}$ and $Rec_{db}$ saturate, but $GF_{db}$ increases and both $IoU_{db}$ and $Pre_{db}$ decline, reflecting over-expansion into voxels far from any data point. Across all cases (Figure 12a-d and Figure 13a-d), this behavior is consistent, with sensitivity primarily set by geometric concavity, branching, curvature, and local point spacing.

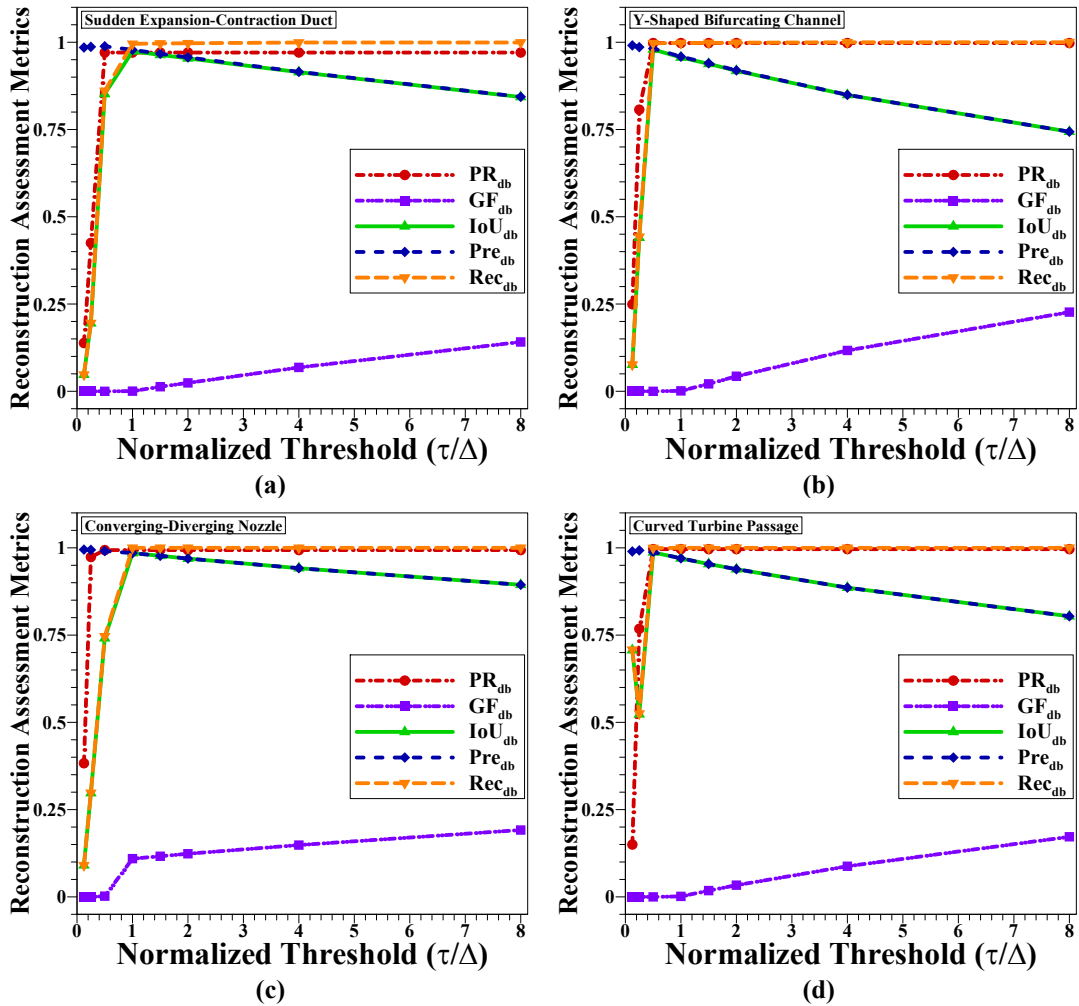

**Figure 12** Effect of the $\tau$ parameter on reconstruction metrics in four geometries of: (a) Sudden expansion–contraction duct, (b) Y-shaped bifurcating channel, (c) Converging–diverging nozzle, and (d) Curved turbine passage.

For the **sudden expansion–contraction duct**, the response to $\frac{\tau}{\Delta}$ is particularly abrupt (Figure 12a). At $\frac{\tau}{\Delta} = 0.125$ and 0.25, the reconstruction remains highly conservative with $PR_{db} = 0.1382$ and 0.4251, while $GF_{db} = 0$ and $Pre_{db} \approx 0.985 - 0.987$. Increasing $\frac{\tau}{\Delta}$ to 0.5 raises $PR_{db}$ to 0.9704, after which it remains essentially unchanged for larger thresholds. The best agreement with the $\alpha$-shape reference occurs around $\frac{\tau}{\Delta} = 1$, where $IoU_{db} = 0.9738$, $Pre_{db} = 0.9782$, and $Rec_{db} = 0.9954$, with $GF_{db}$ still negligible. Beyond this point, enlarging the buffer band primarily increases ghost activation, with $GF_{db}$ rising from 0.0131 at $\frac{\tau}{\Delta} = 1.5$ to 0.1413 at $\frac{\tau}{\Delta} = 8$, accompanied by a monotonic decline in $IoU_{db}$ ($IoU_{db}$ to 0.8431) and $Pre_{db}$ (0.9670 to 0.8436). As shown in Figure 13a, the additional activated voxels concentrate in concave recesses and corner regions that are weakly supported by nearby samples. In the **Y-shaped bifurcating channel**, the $\frac{\tau}{\Delta}$ response is more sensitive because of the strongly concave junction and the thin daughter branches (Figure 12b). At $\frac{\tau}{\Delta} = 0.125$, the reconstruction under-covers the domain with $PR_{db} = 0.2492$, $GF_{db} = 0$, and $Pre_{db} \approx 0.991$. Increasing $\frac{\tau}{\Delta}$ to 0.25 raises $PR_{db}$ to 0.8068, and by $\tau/\Delta = 0.5$ recall saturates at $PR_{db} = 0.9976$. At $\frac{\tau}{\Delta} = 0.5$, $GF_{db}$ remains 0 while $IoU_{db} = 0.9792$, with $Pre_{db} = 0.9820$ and $Rec_{db} = 0.9971$, indicating close agreement with the $\alpha$-shape reference under strict data support. For $\frac{\tau}{\Delta} \geq 1$, $PR_{db}$ stays at 0.9976 and $Rec_{db}$ is essentially 1, but $GF_{db}$ increases from 0.0012 at $\frac{\tau}{\Delta} = 1$ to 0.2269 at $\frac{\tau}{\Delta} = 8$, accompanied by a decline in $IoU_{db}$ (0.9564 to 0.7437) and $Pre_{db}$ (0.9586 to 0.7437). As shown in Figure 13b, the enlarged buffer band progressively fills the concave junction and begins to bridge the inter-branch gap, activating voxels that are weakly supported by nearby points. For the **converging–diverging nozzle**, the response is smoother because the boundary-to-data distance varies gradually along the profile (Figure 12c). Even at $\frac{\tau}{\Delta} = 0.25$, recall is already high with $PR_{db} = 0.9736$, $GF_{db} = 0$, and $Pre_{db} = 0.994$, indicating that a thin buffer band captures most samples. Increasing $\frac{\tau}{\Delta}$ to 0.5 raises $PR_{db}$ to 0.9935 with $GF_{db}$ still very small; however, $IoU_{db} = 0.7412$ and $Rec_{db} = 0.7464$ show that the mask has not yet reached the full reference extent in portions where the $\alpha$-shape boundary sits farther from the sampled locus. The best overlap occurs at $\frac{\tau}{\Delta} = 1$, where $IoU_{db} = Pre_{db} = 0.9848$ and $Rec_{db} = 1$ (Figure 13c), but this comes with a marked increase in ghost activation ($GF_{db} = 0.1095$), consistent with the thicker buffer band admitting voxels that are weakly supported by nearby points. For larger thresholds, $GF_{db}$ increases to 0.1915 at $\frac{\tau}{\Delta} = 8$, while $IoU_{db}$ decreases from 0.9771 at $\frac{\tau}{\Delta} = 1.5$ to 0.8941 at $\frac{\tau}{\Delta} = 8$, with $Pre_{db}$ following the same trend. In this case, $\frac{\tau}{\Delta} \approx 0.5$ emphasizes strict proximity to data, whereas $\frac{\tau}{\Delta} \approx 1$ maximizes agreement with the $\alpha$-shape reference at the cost of additional ghost voxels. For the **curved turbine passage**, which features strong curvature, multiple concavities, and narrow gaps, the sensitivity to $\frac{\tau}{\Delta}$ is again pronounced (Figure 12d). At $\frac{\tau}{\Delta} = 0.125$, the mask yields $PR_{db} = 0.1498$ despite $IoU_{db} = 0.7070$, with $GF_{db} = 0$ and $Pre_{db} = 0.9896$. As shown in Figure 13d, such a thin buffer band can still overlap a substantial fraction of the reference area while excluding many boundary-adjacent samples. At $\frac{\tau}{\Delta} = 0.25$, $PR_{db}$ increases to 0.7679, and by $\frac{\tau}{\Delta} = 0.5$ recall saturates at $PR_{db} = 0.9961$. At this threshold, $GF_{db} = 0$, $IoU_{db} = 0.9870$, $Pre_{db} = 0.9876$, and $Rec_{db} = 0.9994$, indicating near-indistinguishable agreement with the $\alpha$-shape reference under strict data support. For $\frac{\tau}{\Delta} \geq 1$, $PR_{db}$ and $Rec_{db}$ remain saturated, but $GF_{db}$ increases from 0.0014 at $\frac{\tau}{\Delta} = 1$ to 0.1722 at $\frac{\tau}{\Delta} = 8$, while $IoU_{db}$ decreases from 0.9698 to 0.8039 and $Pre_{db}$ follows the same decline. Figure 13d shows that this degradation is driven by the buffer band expanding into concave exterior pockets and into cavities or gaps that are weakly supported by nearby points.

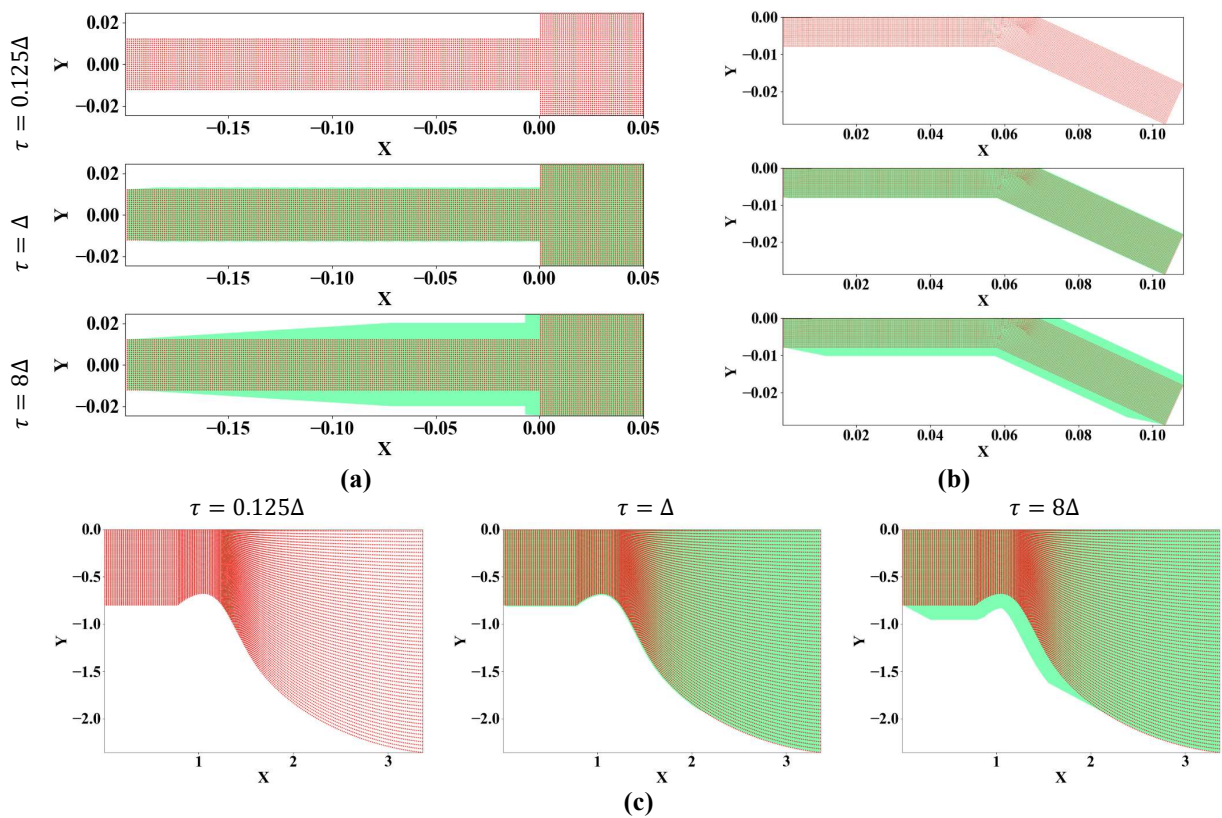

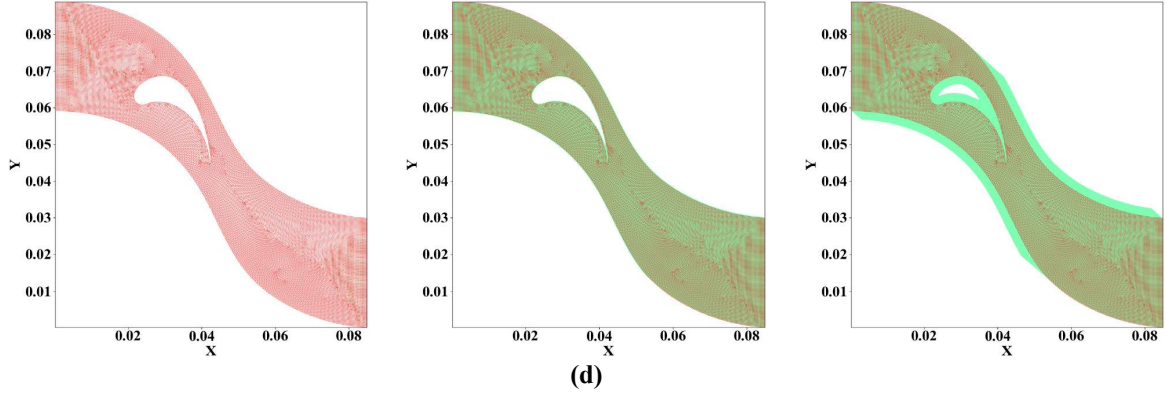

**Figure 13** Distance-based reconstruction outcome for: (a) Sudden expansion–contraction duct, (b) Y-shaped bifurcating channel, (c) Converging–diverging nozzle, and (d-e) Curved turbine passage. The green region denotes the reconstructed domain, and the red points represent the CFD data.

*4.3. Adaptive α-Shape Reconstruction: Effects of β*

In the adaptive $\alpha$-shape reconstruction, $\beta$ enters through the normalized scale in Eq. (27) and defines the circumradius cutoff in Eq. (28) relative to the local resolution $\bar{e}$. Smaller $\beta$ admits larger circumradii, allowing long simplices that can span concavities and yield an inflated, near-convex envelope. Larger $\beta$ tightens the cutoff, retaining only short, well-shaped simplices and potentially oversimplifying the boundary after rasterization. Across all geometries (Figure 14a-d and Figure 15a-d), $PR_{a\alpha}$ and $Rec_{a\alpha}$ remain near 1 for all $\beta$, so the primary $\beta$-dependence is reflected in $GF_{a\alpha}$, $IoU_{a\alpha}$, and $Pre_{a\alpha}$, which measure unsupported activation around the data cloud.

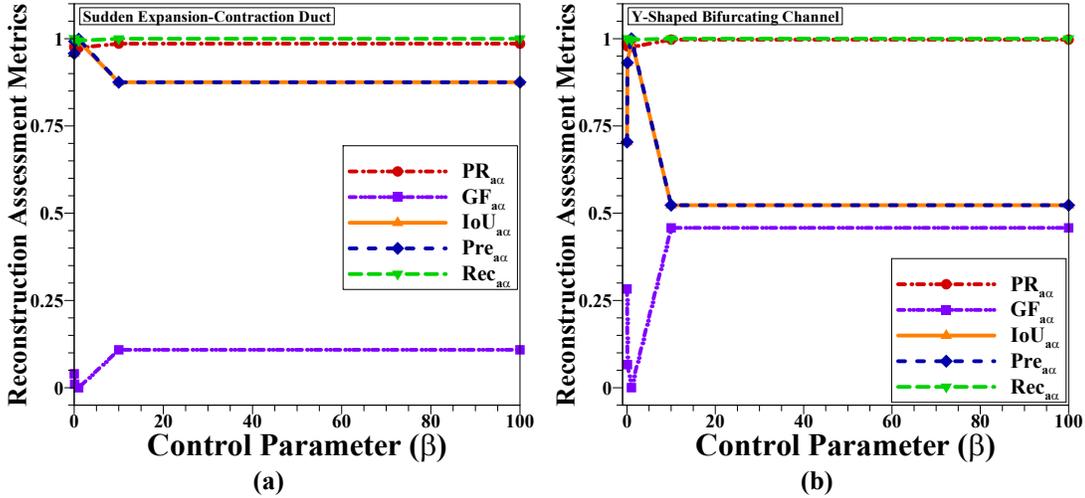

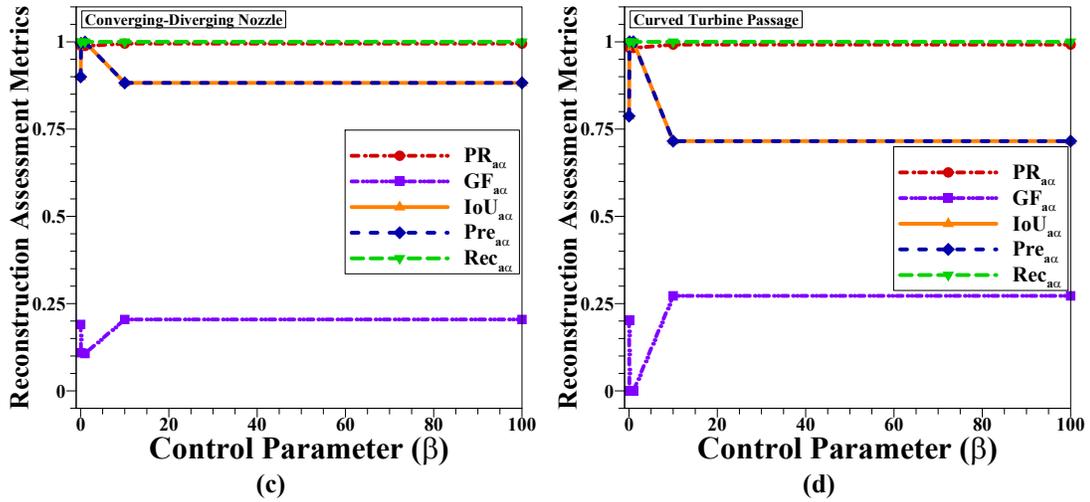

**Figure 14** Effect of the *β* parameter on reconstruction metrics in four geometries of: (a) Sudden expansion–contraction duct, (b) Y-shaped bifurcating channel, (c) Converging–diverging nozzle, and (d) Curved turbine passage.

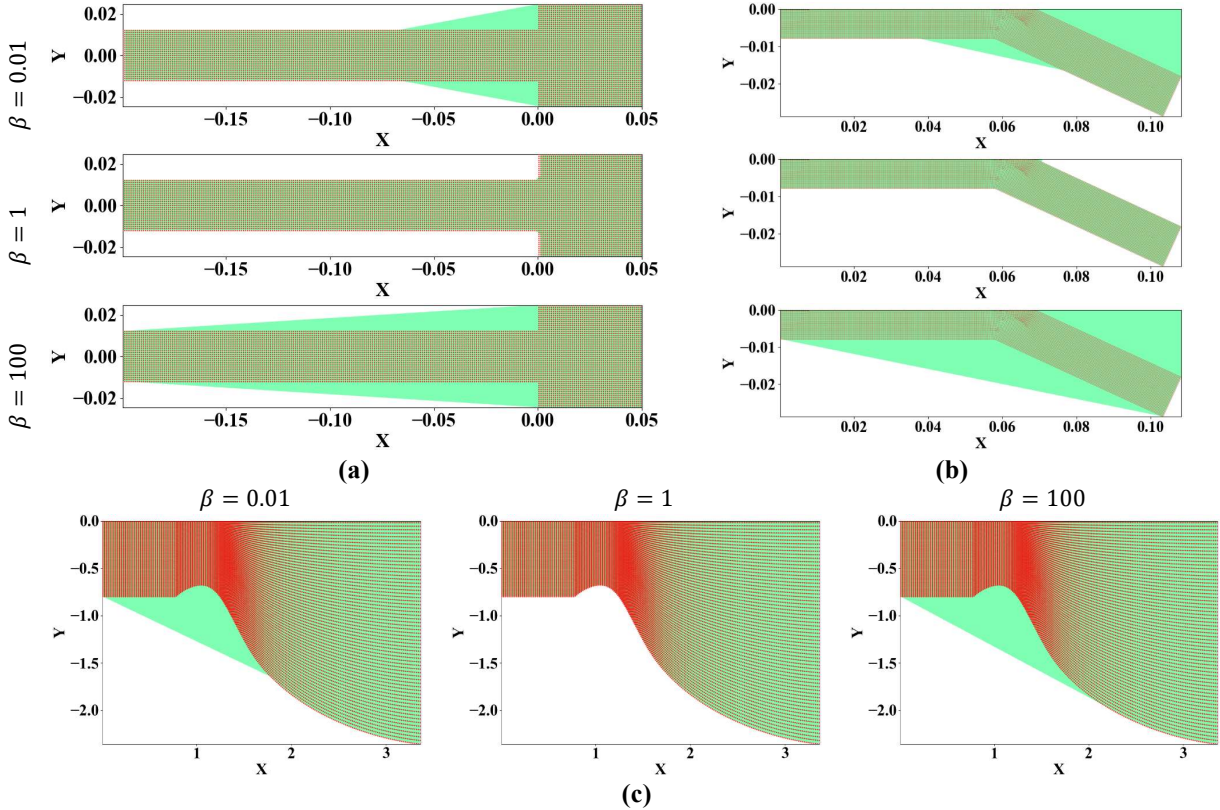

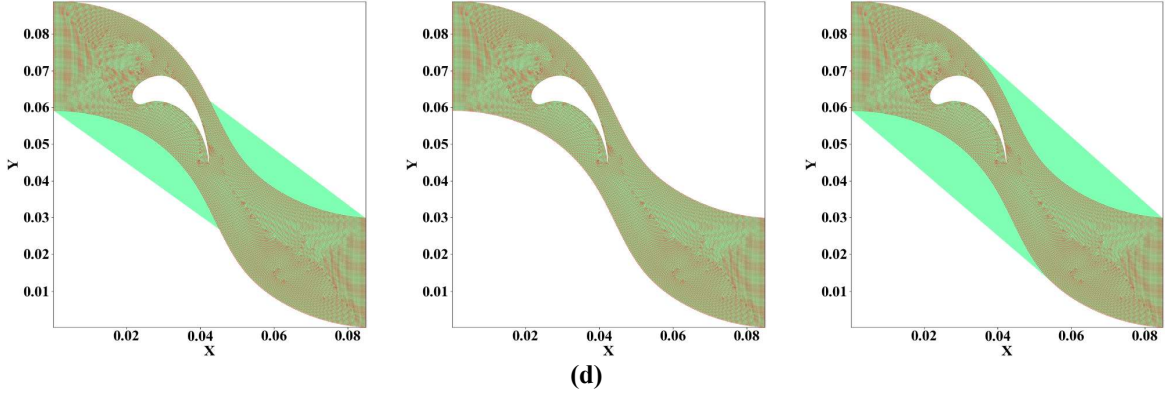

(d)

**Figure 15** Adaptive $\alpha$-shape reconstruction outcome for: (a) Sudden expansion–contraction duct, (b) Y-shaped bifurcating channel, (c) Converging–diverging nozzle, and (d-e) Curved turbine passage. The green region denotes the reconstructed domain, and the red points represent the CFD data.

For the **sudden expansion–contraction duct** (Figure 14a and Figure 15a), $PR_{a\alpha}$ remains stable (about 0.97 to 0.99) for all $\beta$. At $\beta = 0.01$, the permissive circumradius cutoff admits long simplices that span concave gaps, giving mild over-coverage ($GF_{a\alpha} \approx 0.04$, $IoU_{a\alpha} \approx 0.96$). Increasing $\beta$ to 0.1 and 1 tightens the cutoff to the scale of $\bar{e}$, suppressing these spans and driving $GF_{a\alpha}$ to near 0 with $IoU_{a\alpha}$ and $Pre_{a\alpha}$ approaching 1, so the mask matches the reference $\alpha$-shape on the grid. For $\beta \geq 10$, the filter becomes over-restrictive, the boundary simplifies, and unsupported activation increases ($GF_{a\alpha} \approx 0.11$), reducing $IoU_{a\alpha}$ and $Pre_{a\alpha}$ to about 0.88 while still maintaining containment ($Rec_{a\alpha} = 1$). The **Y-shaped bifurcating channel** is the most $\beta$-sensitive case because the bifurcation apex introduces a sharp concavity and the daughter branches are slender (Figure 14b and Figure 15b). At $\beta = 0.01$, the permissive circumradius test admits long simplices that bridge the junction, producing strong over-coverage ($GF_{a\alpha} \approx 0.28$) and low overlap ($IoU_{a\alpha} \approx 0.70$), despite high retention ($PR_{a\alpha} \approx 0.986$, $Rec_{a\alpha} = 1$). Increasing $\beta$ to 0.1 and 1 prunes these bridging simplices, retracting the boundary toward data-supported walls and rapidly suppressing ghost activation ($GF_{a\alpha}: 0.28 \to 0.066 \to {\sim}0$) while improving agreement with the reference ($IoU_{a\alpha}: 0.70 \to 0.93 \to 0.9972$; $Pre_{a\alpha} \to 1$ at $\beta = 1$). For $\beta \geq 10$, the cutoff becomes over-restrictive and the recovered boundary simplifies across the bifurcation, refilling the junction with a coarse enclosure. Containment is preserved ($Rec_{a\alpha} = 1$, $PR_{a\alpha} \approx 0.997$), but unsupported activation increases markedly ($GF_{a\alpha} \approx 0.458$) and $IoU_{a\alpha}$ drops to about 0.52. In the **converging–diverging nozzle** (Figure 14c and Figure 15c), the dependence on $\beta$ is weaker because the boundary curvature and geometric length scales vary smoothly. At $\beta = 0.01$, the permissive cutoff still produces mild over-coverage, with $GF_{a\alpha} \approx 0.19$ and $IoU_{a\alpha} \approx 0.90$. Increasing $\beta$ to 0.1 and 1 tightens the circumradius threshold to the scale of $\bar{e}$ over most of the contour, pruning large simplices and yielding near-unity agreement with the reference (IoU$_{a\alpha}$ up to ~0.999 at $\beta = 1$, with Pre$_{a\alpha}$ and Rec$_{a\alpha}$ essentially 1). For $\beta \geq 10$, the cutoff becomes over-restrictive near local scale changes, and the boundary simplifies after rasterization. This increases unsupported activation (GF$_{a\alpha} \approx 0.20$) and lowers IoU$_{a\alpha}$ and Pre$_{a\alpha}$ to about 0.88, while PR$_{a\alpha}$ and Rec$_{a\alpha}$ remain close to 1. The **curved turbine passage** (Figure 14d and Figure 15d) includes multiple concavities, an internal passage, and narrow gaps, making it highly sensitive to over- and under-filtering. At $\beta = 0.01$, the permissive cutoff admits bridging simplices across internal voids and tight turns, producing an inflated mask (GF$_{a\alpha} \approx 0.20$, IoU$_{a\alpha} \approx 0.79$) while maintaining high retention and containment (PR$_{a\alpha} \approx 0.99$, Rec$_{a\alpha} = 1$). For $\beta = 0.1$ and 1, normalization by $\bar{e}$ suppresses these bridges and yields near-exact agreement with the reference on the Cartesian grid (GF$_{a\alpha} = 0$, IoU$_{a\alpha} = $ Pre$_{a\alpha} = $ Rec$_{a\alpha} = 1$). For $\beta \geq 10$, the cutoff becomes over-restrictive for slender features, the boundary simplifies, and unsupported activation increases (GF$_{a\alpha} \approx 0.27$) with reduced overlap (IoU$_{a\alpha} \approx 0.72$), while PR$_{a\alpha}$ and Rec$_{a\alpha}$ remain high.

### 4.4. Comparison of Mask Generation Runtimes

Table 6 shows a clear method-driven runtime hierarchy, with mask generation time dependent on $\boldsymbol{n} = (n_1, \ldots, n_d)$. The distance-based approach generates masks in $15 - 18\ ms$, indicating a low and nearly constant computational cost. In contrast, both $\alpha$-shape variants require seconds per mask, reflecting a substantially higher processing burden. Within the $\alpha$-shape family, the adaptive formulation consistently improves efficiency, being $1.7 \times$ to $2.6 \times$ faster than the classical $\alpha$-shape for mask generation.

**Table 6** Mask generation runtimes for all geometries using the three reconstruction methods with $\tau = \Delta$, $\alpha = \alpha_{opt}$, and $\beta = 1$.

| Configuration | Mask Generation Time (s) | | |
|---|---|---|---|
| | Distance-Based | Adaptive $\alpha$-Shape | Classical $\alpha$-Shape |
| Sudden expansion–contraction duct | 0.018 | 8.269 | 14.021 |
| Y-shaped bifurcating channel | 0.017 | 5.026 | 9.307 |
| Converging–diverging nozzle | 0.015 | 2.969 | 7.763 |
| Curved turbine passage | 0.016 | 4.054 | 9.313 |

### 4.5. Boundary Inflation for Reduced Edge Misses

In Sections 4.1 to 4.3, the physical domain is reconstructed from scattered CFD datasets using three strategies that yield accurate, practically usable masks across broad parameter ranges. Nevertheless, Figure 13a-d and Figure 15a-d indicate a recurrent voxelization edge effect: a small subset of boundary samples can fall just outside the mask due to sub voxel geometry and grid alignment error, leaving $PR_k$ slightly below 1. A method-agnostic boundary inflation step is therefore applied after reconstruction, dilating the mask by a small expansion factor $\eta$ (for example, $\eta = 1.002$, or 0.2%) to enclose boundary samples (Figure 16a-b). Because excessive dilation activates unsupported regions and increases $GF_k$, $\eta$ is kept minimal to drive $PR_k$ toward 1 while maintaining a negligible change in $GF_k$.

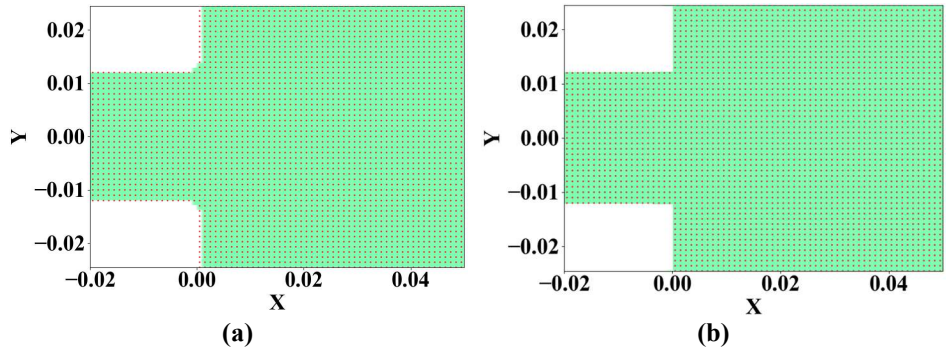

**Figure 16** Boundary inflation refinement for mitigating edge under-coverage: (a) original mask, (b) inflated mask. The green region denotes the reconstructed domain of first geometry, and the red points represent the CFD data.

**Table 7** Effect of boundary inflation on $PR_k$ and $GF_k$ for distance-based and adaptive $\alpha$-shape reconstructions.

| Configuration | Original | | Inflated | | Original | | Inflated | |
|---|---|---|---|---|---|---|---|---|
| | $PR_{db}$ | $GF_{db}$ | $PR_{db}$ | $GF_{db}$ | $PR_{a\alpha}$ | $GF_{a\alpha}$ | $PR_{a\alpha}$ | $GF_{a\alpha}$ |
| Sudden expansion–contraction duct | 0.9704 | 0.0004 | 1 | 0.0010 | 0.9741 | 0 | 1 | 0.0002 |
| Y-shaped bifurcating channel | 0.9976 | 0.0012 | 0.9994 | 0.0015 | 0.9761 | 0.0003 | 0.9913 | 0.0005 |
| Converging–diverging nozzle | 0.9935 | 0.1095 | 0.9978 | 0.1100 | 0.9881 | 0.1071 | 0.9951 | 0.1074 |
| Curved turbine passage | 0.9961 | 0.0014 | 0.9998 | 0.0017 | 0.9826 | 0 | 0.9964 | 0.0008 |

Table 7 quantifies the effect of boundary inflation and shows a consistent trend across all geometries. For the distance-based method, $PR_{db}$ increases by 3.05% in the sudden expansion–contraction duct, 0.18% in the Y-shaped channel, 0.43% in the converging–diverging nozzle, and 0.37% in the curved turbine passage. Similar gains occur for the adaptive $\alpha$-shape, with $PR_{a\alpha}$ rising by 2.66%, 1.56%, 0.71%, and 1.40% across the same cases. These improvements are accompanied by only negligible changes in ghost content: $GF_{db}$ increases by 0.06%, 0.03%, 0.05%, and 0.03%, and $GF_{a\alpha}$ increases by 0.02%, 0.02%, 0.03%, and 0.08%. After inflation, both approaches converge to nearly identical $PR_k$ and $GF_k$ across all cases, yielding practically indistinguishable reconstructed domains and boundaries, as illustrated in Figure 17a-d.

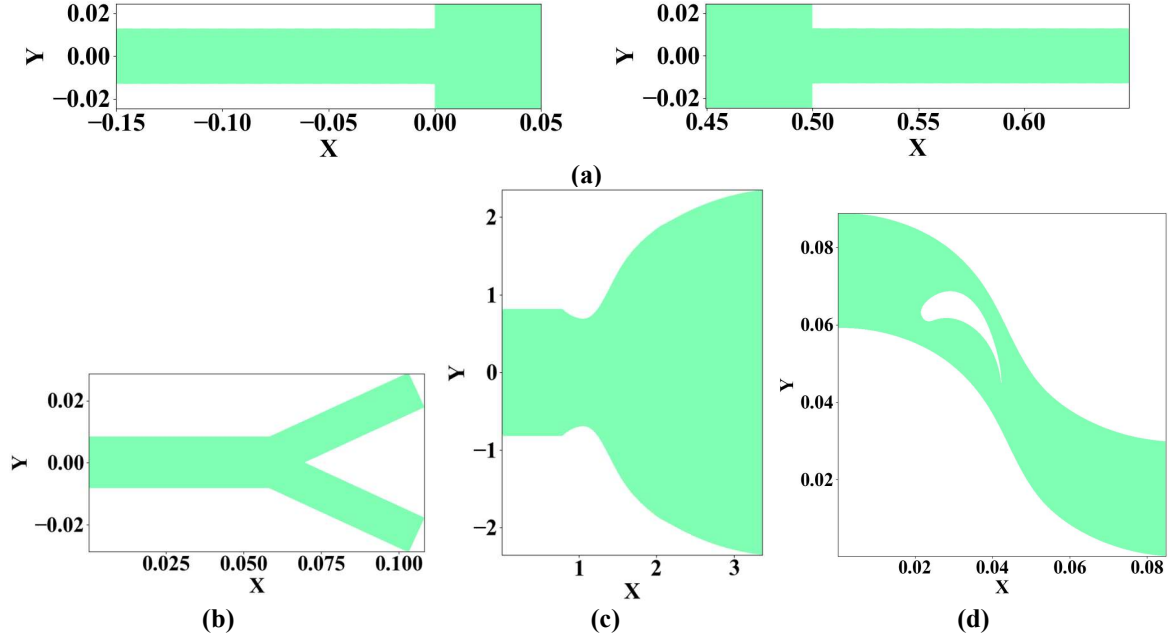

**Figure 17** Final domain reconstruction and boundary recovery for (a) Sudden expansion–contraction duct, (b) Y-shaped bifurcating channel, (c) Converging–diverging nozzle, and (d) Curved turbine passage.

## 5. Conclusion

Interpolating scattered CFD datasets onto a uniform Cartesian grid can produce a convex-hull type envelope and activate nonphysical regions, so a dedicated domain reconstruction step is required before exporting CNN-ready fields. Accordingly, a reconstruction framework is developed and evaluated using three distinct strategies: novel distance-based masking with morphological closing, classical $\alpha$-shape boundary recovery, and a novel adaptive $\alpha$-shape formulation that normalizes $\alpha$ using the local data resolution. To enable objective evaluation beyond visual inspection, a set of quantitative, topology-aware evaluation metrics is introduced for the first time to measure data retention, suppression of unsupported regions, overlap consistency, and connectivity of the recovered domain.

Parameter robustness shows a clear hierarchy. The distance-based strategy operates reliably with a single universal choice, $\tau = \min_{1 \leq i \leq d \in \{2,3\}}(\Delta x_i)$, across all tested geometries, and the adaptive $\alpha$-shape method remains stable with $\beta = 1$ without case-specific retuning. In contrast, the classical $\alpha$-shape formulation requires geometry-dependent tuning of $\alpha$ parameter to balance boundary under-coverage against inclusion of unsupported regions. For instance, the optimal $\alpha$ values are 100 for the sudden expansion–contraction duct, 1000 for the Y-shaped bifurcating channel, 10 for the converging–diverging nozzle, and $100 - 150$ for the curved turbine passage, highlighting its sensitivity to scaling and sampling variations.

Computational cost follows the same trend. Notably, the distance-based approach is nearly constant across the tested geometries, requiring only about 15 to 18 $ms$ to generate the mask in every configuration, while remaining orders of magnitude faster overall. Relative to the classical $\alpha$-shape baseline, the distance-based method is approximately 500 to 800 times faster, whereas the adaptive $\alpha$-shape technique reduces runtime by about $1.7 \times$ to $2.6 \times$, staying consistently faster than the classical variant but slower than the distance-based approach.

Boundary inflation provides a lightweight, method-independent refinement that corrects a voxelization edge effect in which some boundary-adjacent samples fall marginally outside an otherwise accurate mask. With a minimal, controlled dilation of $\eta = 0.2\%$, the retained-sample fraction increases by 0.18% to 2.96% for the distance-based masks and by 0.70% to 2.59% for the adaptive $\alpha$-shape method, driving retention to near unity across all cases. This improvement incurs only a negligible increase in unsupported region activation, rising by 0.03% to 0.06% for the distance-based method and by at most 0.08% for the adaptive $\alpha$-shape method, indicating that the refinement recovers missed boundary samples without meaningful expansion into nonphysical regions.

Overall, the distance-based method is the preferred default, providing the most reliable trade-off between accuracy, robustness, minimal parameter tuning, and computational efficiency. Adaptive $\alpha$-shape approach remains a strong alternative when the CFD grid-size information required for setting the distance-based threshold is unavailable. Although the present study examines only four geometries, a dedicated web application delivers the workflow end to end, enabling users to upload 2D ASCII datasets for any geometry, adjust method parameters, and automatically reconstruct boundaries and export CNN-ready outputs.

**Appendix: Additional $\alpha$ Parameter Study**

For the **sudden expansion–contraction duct** (Figure 18a and Figure 19a), $PR_\alpha$ remains high and nearly constant from $\alpha = 0.01$ to 1000, decreasing only slightly from about 0.9845 to 0.9726. Over the same range, $GF_\alpha$ decreases monotonically from roughly 0.109 to 0 by $\alpha = 500$, while $AVF_\alpha$ contracts from about 0.88 to 0.77. The reconstruction stays topologically connected with $N_{c,\alpha} = 1$ throughout this interval. A failure occurs only at $\alpha = 1430$, where $PR_\alpha$ collapses to about 0.34, $AVF_\alpha$ drops to 0.26, and $N_{c,\alpha}$ increases to 2, indicating fragmentation induced by overly restrictive circumradius filtering. For the **Y-shaped bifurcating channel** (Figure 18b and Figure 19b), $PR_\alpha$ remains essentially 1 up to $\alpha = 1$, then decreases gradually to about 0.976 for $100 \leq \alpha \leq 4500$, before collapsing to 0.044 at $\alpha = 5000$. $GF_\alpha$ is very high and nearly flat for small $\alpha$, around 0.458 (with a slight increase at $\alpha = 10$), then drops sharply to about 0.074 at $\alpha = 100$ and becomes nearly 0 for $\alpha \geq 500$. In parallel, $AVF_\alpha$ contracts from about 0.64 at $\alpha = 0.01$ to roughly 0.34 by $\alpha = 500$, after which it plateaus until the final collapse. The reconstruction remains connected with $N_{c,\alpha} = 1$ throughout the stable range, then fragments abruptly at $\alpha = 5000$ where $N_{c,\alpha}$ increases to 19. This strong sensitivity is consistent with the junction's pronounced concavity and the presence of thin branches: permissive $\alpha$ values promote circumradius-supported bridging across the junction, whereas overly strict filtering at large $\alpha$ truncates the slender branch segments and induces fragmentation. For the **converging–diverging nozzle** (Figure 18c and Figure 19c), $PR_\alpha$ decreases only slightly from 0.991 to 0.984 up to $\alpha = 45$, then drops abruptly at $\alpha = 50$ and continues falling to 0.033 by $\alpha = 125$. $GF_\alpha$ decreases from about 0.204 at $\alpha = 0.01$ to roughly 0.107 over $10 \leq \alpha \leq 45$, and reaches 0 by $\alpha = 75$. Similarly, $AVF_\alpha$ contracts smoothly from about 0.71 to 0.62 up to $\alpha = 45$, then collapses rapidly after $\alpha = 50$ to below 0.01 by $\alpha = 125$. The reconstruction remains connected with $N_{c,\alpha} = 1$ up to $\alpha = 75$, after which fragmentation emerges with $N_{c,\alpha} = 2$ at $\alpha = 100$ and 7 at $\alpha = 125$. The sharp transition near $\alpha \approx 50$ indicates a threshold beyond which circumradius filtering begins to remove simplices needed to represent the locally narrowest or most sparsely supported boundary segments, producing rapid under-coverage and subsequent fragmentation.

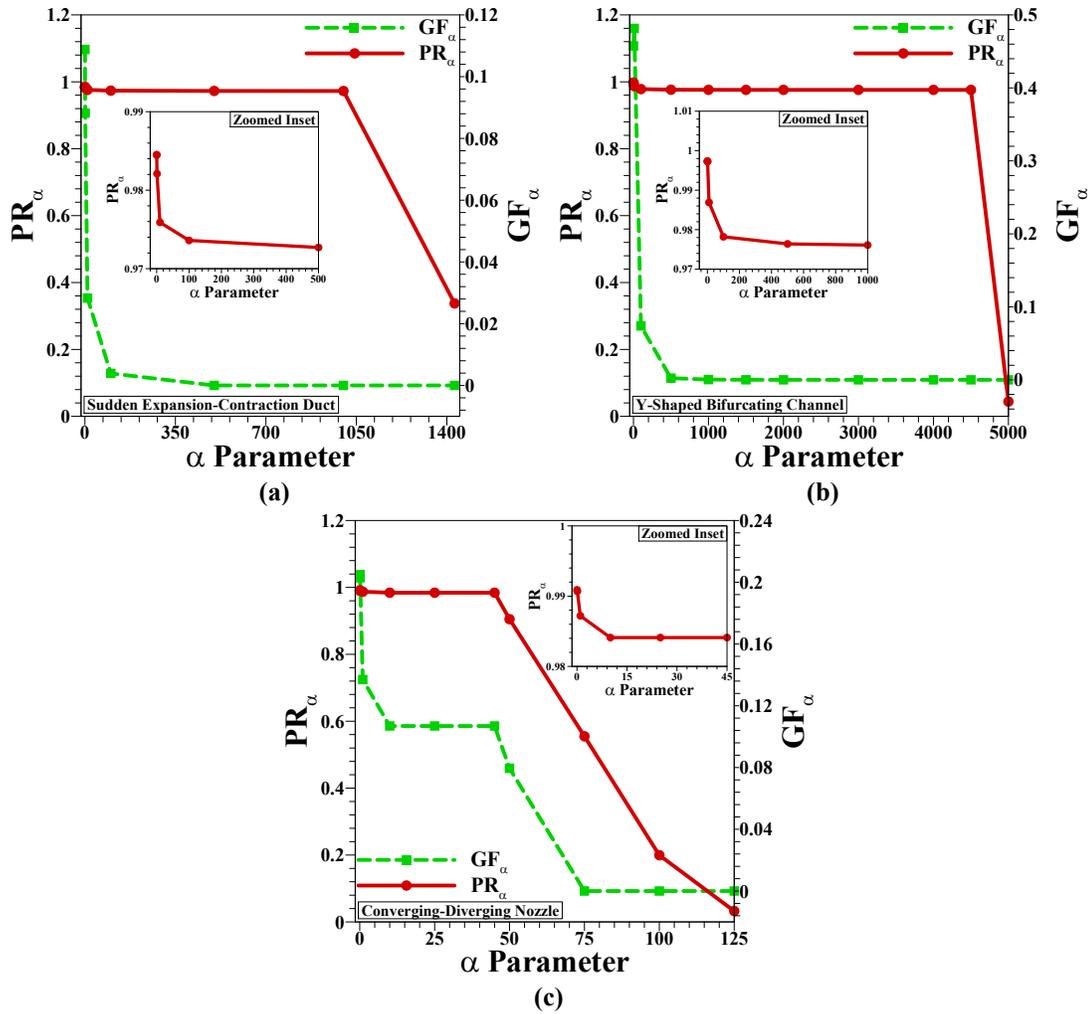

**Figure 18** Effect of the $\alpha$ parameter on $PR_\alpha$ and $GF_\alpha$ in three geometries of: (a) Sudden expansion–contraction duct, (b) Y-shaped bifurcating channel, and (c) Converging–diverging nozzle.

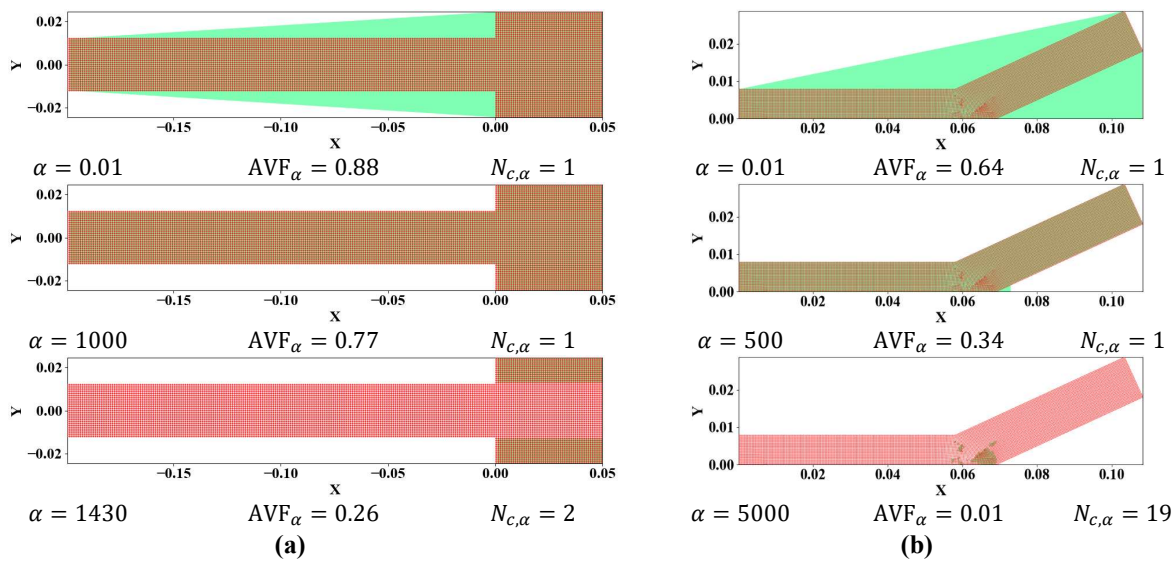

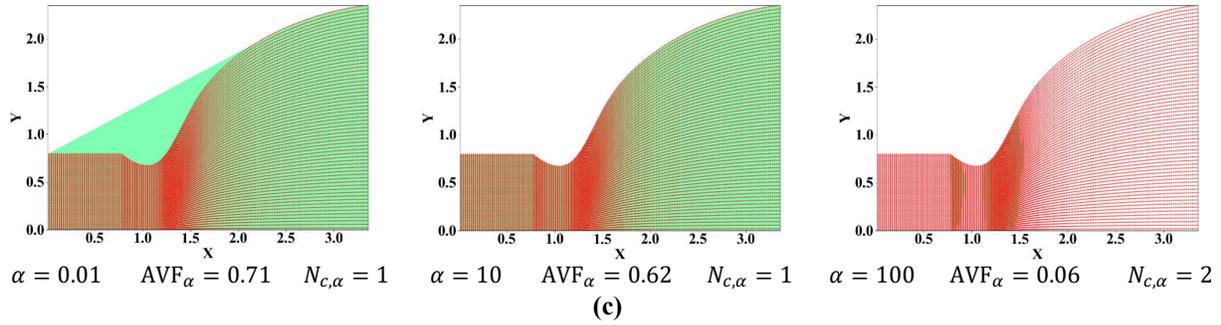

**Figure 19** $\alpha$-shape reconstruction outcome for: (a) Sudden expansion–contraction duct, (b) Y-shaped bifurcating channel, and (c) Converging–diverging nozzle. The green region denotes the reconstructed domain, and the red points represent the CFD data.

**Declaration of Interests**

The authors declare that they have no known competing financial interests or personal relationships that could have appeared to influence the work reported in this paper.

**Data and Code Availability**

Data will be made available by the corresponding author upon reasonable request. In addition, the integrated web application is available at https://CFDtoCNN.streamlit.app, and a brief tutorial is provided in the Supplementary Material.

**Acknowledgements**

This work is part of the project MoReDigi (Desarrollo de modelos reducidos basados en la física para la digitalización industrial) funded by the Government of the Basque Country through research grants ELKARTEK (KK-2024/00117).

**Funding Statement**

This work is part of the project MoReDigi (Desarrollo de modelos reducidos basados en la física para la digitalización industrial) funded by the Government of the Basque Country through research grants ELKARTEK (KK-2024/00117).

**CRediT Authorship Contribution Statement**

**Mehran Sharifi:** Writing-review & editing, Writing-original draft, Visualization, Validation, Software, Resources, Methodology, Investigation, Formal Analysis, Data Curation, Conceptualization.

**Gorka S. Larraona:** Writing-review & editing, Supervision, Resources, Project administration, Methodology, Investigation, Formal Analysis, Data Curation, Conceptualization, Funding Acquisition.

**Alejandro Rivas:** Writing-review & editing, Supervision, Resources, Project administration, Methodology, Investigation, Formal Analysis, Data Curation, Conceptualization.